\documentclass[preprint,12pt]{elsarticle}

\usepackage{amssymb}

\makeatletter
\def\ps@pprintTitle{%
  \let\@oddhead\@empty
  \let\@evenhead\@empty
  \let\@oddfoot\@empty
  \let\@evenfoot\@oddfoot
}
\makeatother

\begin{document}

\begin{frontmatter}



\title{ Prediction of COVID-19 by Its Variants using Multivariate Data-driven Deep Learning Models}


\author[inst1]{Akhmad Dimitri Baihaqi}

\affiliation[inst1]{organization={Department of Informatics, Faculty of Computer Science, Brawijaya University},
            addressline={Veteran 8}, 
            city={Malang},
            postcode={65145}, 
            state={East Java},
            country={Indonesia}}

\author[inst1]{Novanto Yudistira}
\author[inst1]{Edy Santoso}

\begin{abstract}
The Coronavirus Disease 2019 or the COVID-19 pandemic has swept almost all parts of the world since the first case was found in Wuhan, China, in December 2019. With the increasing number of COVID-19 cases in the world, SARS-CoV-2 has mutated into various variants. Given the increasingly dangerous conditions of the pandemic, it is crucial to know when the pandemic will stop by predicting confirmed cases of COVID-19. Therefore, many studies have raised COVID-19 as a case study to overcome the ongoing pandemic using the Deep Learning method, namely LSTM, with reasonably accurate results and small error values. LSTM training is used to predict confirmed cases of COVID-19 based on variants that have been identified using ECDC's COVID-19 dataset containing confirmed cases of COVID-19 that have been identified from 30 countries in Europe. Tests were conducted using the LSTM and BiLSTM models with the addition of RNN as comparisons on hidden size and layer size. The obtained result showed that in testing hidden sizes 25, 50, 75 to 100, the RNN model provided better results, with the minimum MSE value of 0.01 and the RMSE value of 0.012 for B.1.427/B.1.429 variant with hidden size 100. In further testing of layer sizes 2, 3, 4, and 5, the result shows that the BiLSTM model provided better results, with minimum MSE value of 0.01 and the RMSE of 0.01 for the B.1.427/B.1.429 variant with hidden size 100 and layer size 2.
\end{abstract}




\end{frontmatter}


\section{Introduction}
\label{sec:sample1}
The Coronavirus Disease 2019, or COVID-19 pandemic, has swept almost all countries worldwide since the first case was found in Wuhan, China, in December 2019. COVID-19 was carried by the Novel Coronavirus (SARS-CoV-2), which was first identified on January 7th, 2020 \cite{Gorbalenya2020}. The virus could be spread through water droplets and direct contacts, such as coughing and sneezing. Since COIVID-19 was first identified, WHO declared it a global pandemic on March 11th, 2020. COVID-19 has spread to 220 countries, with 221,134,742 confirmed cases and 4,574,089 deaths \cite{WHO2021}.
With the increasing number of COVID-19 cases in the world, SARS-CoV-2 has mutated into various variants. Since the first variant, Alpha (B.1.1.7), was discovered in the UK until February 18th, 2022, 12 variants of COVID-19 have been identified by WHO, with the Omicron and Delta variants now spreading rapidly \cite{WHO2022}. Given the increasingly dangerous conditions of the pandemic, it is crucial to know when the pandemic will stop by predicting confirmed cases of COVID-19. The prediction information can also serve as an early warning to all elements of society if the number of cases has reached an alarming level; therefore, many studies have raised COVID-19 as a case study to overcome the ongoing pandemic. The approach initially used traditional Machine Learning models to get better statistical results \cite{Shahid2020}. One uses the Support Vector Regression (SVR) model to predict the total positive cases of COVID-19, daily new cases, death cases, and daily death cases \cite{Parbat2020}. However, the results could be more accurate and produce a high error value compared to the Deep Learning method.

So based on these results, Deep Learning is the right choice, especially for analysis and prediction since COVID-19 has data that consists of values in a specific time (time series data). So the correct prediction is time series prediction, and the suitable model to be used is Long Short Term Memory (LSTM).

LSTM has been widely used, especially for COVID-19 cases that are spreading rapidly today. On a global scale, LSTM has been used to predict the growth of confirmed cases of COVID-19 evaluated using RMSE \cite{Yudistira2020}\cite{Yudistira2021} and also to predict positive cases. Death cases of COVID-19 in 10 major countries with MAE and RMSE evaluations \cite{Shahid2020}. While on a country scale, the use of LSTM to predict time series was carried out in Canada for the spread of COVID-19 \cite{Chimmula2020} and Russia, Peru, and Iran for positive cases of COVID-19 \cite{Wang2020}. Based on the implementation of the LSTM model, which is widely used to predict time series, in this study, researchers will use the LSTM model to predict the time series of the number of confirmed cases of COVID-19 for each identified variant.

\section{Methods}
\subsection{COVID-19 Dataset}
The dataset used in the study belongs to the European Center for Disease Prevention and Control (ECDC), which contains the number of confirmed COVID-19 cases per identified variant from 30 countries in Europe. 
\begin{table}[ht]
	\centering
	\caption{Variants Information \cite{WHO2022}}
	\label{table11}
	\begin{tabular}{@{}cccc@{}}
	\hline Variant (Pango Lineage)& WHO Label& Origin\\
    \hline
B.1.1.7& Alpha& United Kingdom\\
B.1.351& Beta& South Africa\\
B.1.427/B.1.429& Epsilon& South Africa\\
B.1.525& Eta& United Kingdom and Nigeria\\
B.1.616& -& France \cite{Fillatre2021}\\
B.1.617.1& Kappa& India\\
B.1.617.2& Delta& India\\
B.1.620& -& Cameroon, Africa \cite{Dudas2021}\\
B.1.621& Mu& South America\\
BA.1& Omicron& South Africa\\
BA.2& Omicron& South Africa\\
BA.2.75& Omicron& India\\
BA.4& Omicron& South Africa\\
BA.5& Omicron& South Africa\\
BQ.1& Omicron& Nigeria\\
C.37& Lambda& Peru, Chile, USA and Germany\\
Other& -&-\\
P.1& Gamma& Brazil \cite{Faria2021}\\
P.3& -&Philippines \cite{Francis2021}\\
UNK& Unkown& Unknown\\
XBB& Omicron& South Asia\\
    \hline
	\end{tabular}
\end{table}
The source of this data comes from the Global Initiative on Sharing Avian Influenza Data (GISAID) and The European Surveillance System (TESSy), recorded on weekly and continuously updated to the latest date. The dataset we use started from 2020 Week 1 to 2022 Week 49 and contains 21 COVID-19 variants. The naming of those variants is based on the nomenclature of the phylogenic framework \cite{Rambaut2020} or commonly known, as Pango Nomenclature. 21 variants we use in this study are shown in Table \ref{table11}. The dataset was filtered using only the data sourced from GISAID because it has earlier records than the TESSy one. Also, all variants in Table \ref{table11} are sourced from GISAID. The dataset snippets for variant B.1.1.7 are shown in Table \ref{table1} for the five first countries and their confirmed case.

\begin{table}[ht]
	\centering
	\caption{5 First Countries for variant B.1.1.7}
	\label{table1}
	\begin{tabular}{@{}cccc@{}}
	\hline Country& Year& Variant& Number Detection \\
    \hline
		Austria & 2021-01 & B.1.1.7 & 56 \\
		Belgium & 2021-01 & B.1.1.7 & 65 \\
		Bulgaria & 2021-01 & B.1.1.7 & 20 \\
		Croatia & 2021-01 & B.1.1.7 & 0 \\
		Cyprus & 2021-01 & B.1.1.7 & 1 \\
    \hline
	\end{tabular}
\end{table}

First, the dataset is divided based on the variant, then divided into training and testing data with a proportion of 80:20, by dividing based on the unit of time, which is week. There are 149 weeks for each variant, so the training data will contain the first 123 weeks, while the testing data will contain the last 26 weeks. 

\subsection{LSTM}
LSTM, or Long Short Term Memory, is one type of RNN or Recurrent Neural Network [10], is also a type of strong artificial neural network suitable for solving problems from time series data. The LSTM architecture consists of a cell state and three gates: input, forget, and output.

Cell state functions to channel information from one sequence to another with changes to the value will be passed through 3 gates. Forget gate serves to determine the current input and the previous hidden state that will be stored or discarded with the output in the form of forget gate value (ft). The input gate updates the cell state's value by calculating the candidate cell state (Ct) and input gate (it). Furthermore, the output gate calculates the value of the hidden state \cite{Yu2019}. The architecture of LSTM can be seen in Figure \ref{fig1} below.

\begin{figure}[ht]
	\centering
\includegraphics[width=1.0\textwidth]{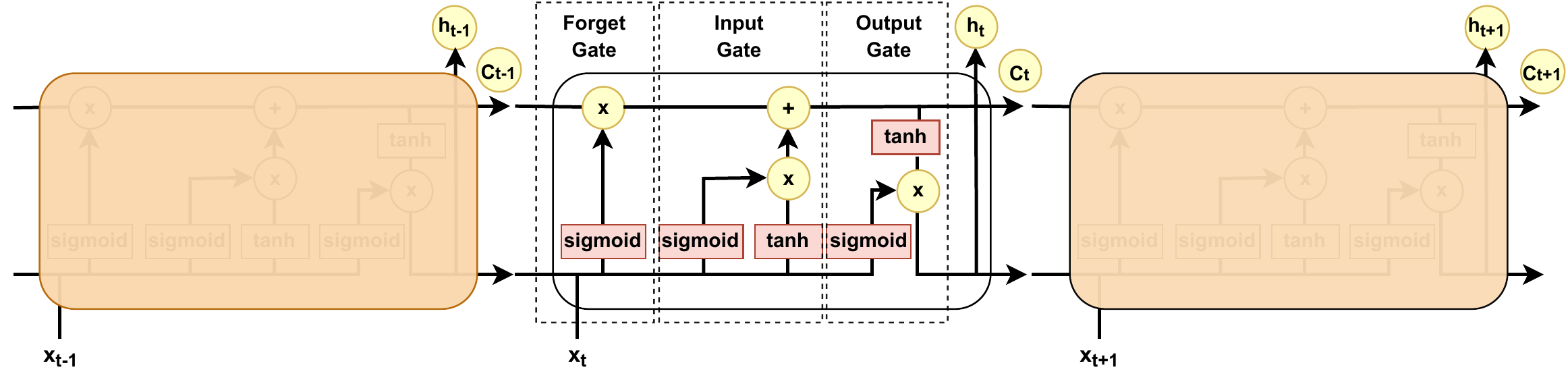}
	\caption{Architecture of LSTM consists of forget gate, input gate and output gate.}
	\label{fig1}
\end{figure}

The formula for the input gate can be seen in Equation \ref{Eq2}, forget gate in Equation \ref{Eq3}, and output gate in Equation \ref{Eq6}.

\begin{eqnarray}
\ i_{t}=\sigma(w_{f}*[h_{(t-1)},x_{t} ]+ b_{i})\label{Eq2}
 \end{eqnarray}
 \begin{eqnarray}
    \ f_{t}=\sigma(w_{i}*[h_{(t-1)},x_{t} ]+ b_{f})
	\label{Eq3}
 \end{eqnarray}
 \begin{eqnarray}
    \ g_{t}=tanh(w_{c}*[h_{(t-1)},x_{t} ]+ b_{c})
	\label{Eq4}
 \end{eqnarray}
 \begin{eqnarray}
    \ c_{t}=f_{t}*C_{(t-1)}+i_{t}*g_{t})
	\label{Eq5}
 \end{eqnarray}
 \begin{eqnarray}
     \ o_{t}=\sigma(w_{o}*[h_{(t-1)},x_{t} ]+ b_{o})
	\label{Eq6}
 \end{eqnarray}
 \begin{eqnarray}
     \ h_{t}=O_{t}*tanh(C_{t})
	\label{Eq7}
\end{eqnarray}

\subsection{BiLSTM}
Bidirectional LSTM or BiLSTM is a development model of conventional LSTM, consisting of two LSTMs in each layer one is used for forwarding propagation, and the other is  used for backward propagation \cite{Siami2019}. 
\begin{figure}[ht]
	\centering
	\includegraphics[width=1.0\textwidth]{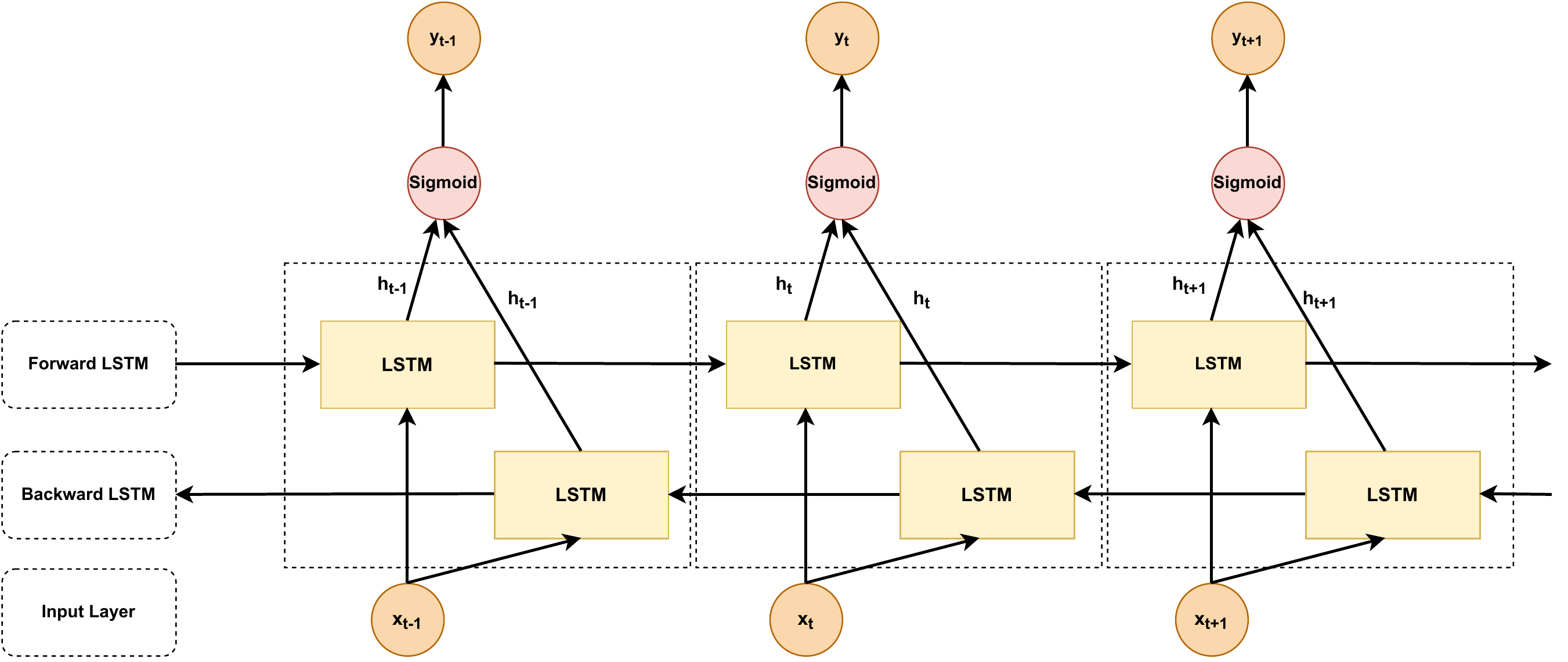}
	\caption{Architecture of BiLSTM.}
	\label{fig2}
\end{figure}
Bi-LSTM can improve the model's ability to learn during the training process and remember long sequences. In addition, BiLSTM is used to obtain minor prediction errors and high accuracy \cite{Shahid2020}. The architecture of BiLSTM can be seen in Figure \ref{fig2}. Each LSTM box is an LSTM cell with architecture according to Figure \ref{fig1}.

\subsection{Robust Scaler}
Before being used, the dataset was normalized using Robust Scaler. It is necessary so that when data is entered into the model, it does not give bad output due to incomplete data or the presence of an outlier. This method was chosen because the dataset used in this research contained outliers. As we know, the nature of the COVID-19 case dataset has many unexpected spikes in numbers. So that causes outliers to exist in the dataset. The scale function is defined in the following as Equation \ref{Eq1}
\begin{eqnarray}
	\chi^{'}=(\chi_{i}-\chi_{median})/(\chi_{75}-\chi_{25})
	\label{Eq1}
\end{eqnarray}

\begin{figure}[ht]
	\centering
\includegraphics[width=0.9\textwidth]{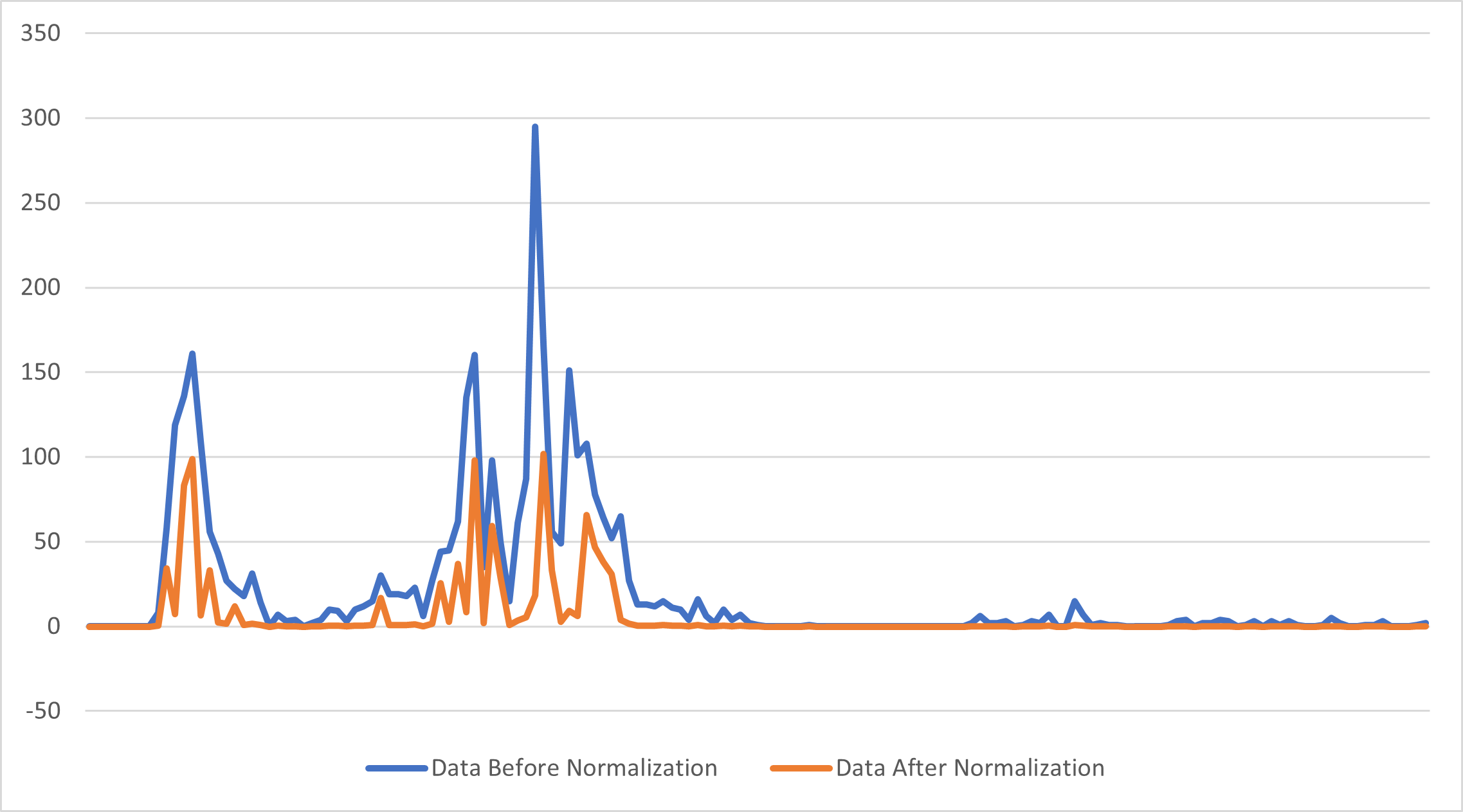}
	\caption{Comparison using Robust Scaler.}
	\label{fig3}
\end{figure}

Where $x_i$ is the input data and $x'$ is the normalized output, while $x75$ and $x25$ are the interquartile range (IQR) which is the range between quartile 1 ($25^th$ quantile) and quartile 3 ($75^th$ quantile) of the data. Here we show the applied Robust Scaler to our dataset in Figure \ref{fig3}, how before and after Robust Scaler was applied to the dataset used in this research.

Outliers from the data still exist, but the difference in value between the data and the outliers could be more negligible. Thus, the shape of the data is maintained, and the value of each data is more evenly distributed.

\subsection{Adam Optimizer}
Optimization is widely used to obtain the highest possible accuracy with minimum training time. However, general optimization methods such as stochastic gradient-based optimization still need to be more effective in high-dimensional parameter spaces. Thus, it is necessary to use a more efficient optimization, namely Adam Optimizer. Adam optimizer has several advantages, including small memory usage and an adaptive learning rate for each parameter \cite{Kingma2015}.
The algorithm of Adam Optimizer can be seen in Equation \ref{Eq10} to Equation \ref{Eq16}.

\begin{eqnarray}
\label{Eq10}
\ g_{t}\leftarrow \nabla_{\theta}f_{t}(\theta_{(t-1)})
\end{eqnarray}
\begin{eqnarray}
\label{Eq11}
\ m_{t}\leftarrow \beta_{1}\cdot m_{(t-1)}+(1-\beta_{1})
\end{eqnarray}
\begin{eqnarray}
\label{Eq12}
\ v_{t}\leftarrow \beta_{2}\cdot v_{(t-1)}+(1-\beta_{2})
\end{eqnarray}
\begin{eqnarray}
\label{Eq13}
\ \hat{m}_{t}\leftarrow m_{t}⁄(1-\beta_{1}^{t})
\end{eqnarray}
\begin{eqnarray}
\label{Eq14}
\ v_{t}\leftarrow v_{t}⁄(1-\beta_{2}^{t})
\end{eqnarray}
\begin{eqnarray}
\label{Eq15}
\ \theta_{t}\leftarrow \theta_{(t-1)}-\alpha\cdot \hat{m}_{t}⁄(\sqrt{(\hat{v}_{t}+\epsilon)})
\label{Eq16}
\end{eqnarray}

\subsection{MSE and RMSE}
In this study, each model's performance is measured using Mean Squared Error (MSE) and Root Mean Squared Error (RMSE). Both methods are differentiable functions, meaning it is easier to perform mathematical computations than non-differentiable functions like MAE. MSE is defined as the average square of the difference between the estimated and actual value, and RMSE is the root of MSE \cite{Chai2014}. The formula for MSE is shown in Equation \ref{Eq8} and RMSE in Equation \ref{Eq9}.
\begin{eqnarray}
	\ MSE = 1/N * \Sigma^{N}_{i=1}(y_{i}-y^{'}_{i})
	\label{Eq8}
 \end{eqnarray}
 
\begin{eqnarray}
	\ RMSE = \sqrt{1/N * \Sigma^{N}_{i=1}(y_{i}-y^{'}_{i})}
	\label{Eq9}
 \end{eqnarray}
 
\section{Result and Discussion}
This paper aims to find the optimal model configuration for predicting COVID-19 confirmed cases based on its variant from 30 countries across Europe. LSTM and BiLSTM were implemented as a model for prediction along with RNN for comparison. The test was done in 2 types, Univariate and Multivariate. Both were going through 2 test stages, hidden size test and layer size test. Similar research on testing the hidden and layer size on LSTM has been done before \cite{Yudistira2020}, but we take a different approach. 

Our first test was conducted by testing the number of hidden sizes implemented, combined with layer size 1 for each test, to avoid bias. Hidden sizes tested are 25, 50, 75, and 100. The optimal hidden size was then obtained by examining which hidden size gave the largest frequency of minimum loss value from each COVID-19 variant. Then the second test was carried out by testing the number of layer sizes implemented starting from 2, 3, 4, to 5, combined with the optimal hidden size value obtained from the previous test. The result of the layer size test was determined in the same approach as the hidden size one. The final result is the optimal model configuration of both hidden and layer sizes to be implemented.

The training process of each model is performed for 1000 epochs using a sequence length of 10 and an optimizer value of 0.01. The trained model is then used in testing. Each Univariate and Multivariate test is performed individually and not continuously from one another, and the same applies to hidden size and layer size tests. Loss is then calculated to determine the performance of each configuration and model.

\subsection{Univariate Test}
\begin{figure}[p]
	\centering	\includegraphics[width=0.9\textwidth]{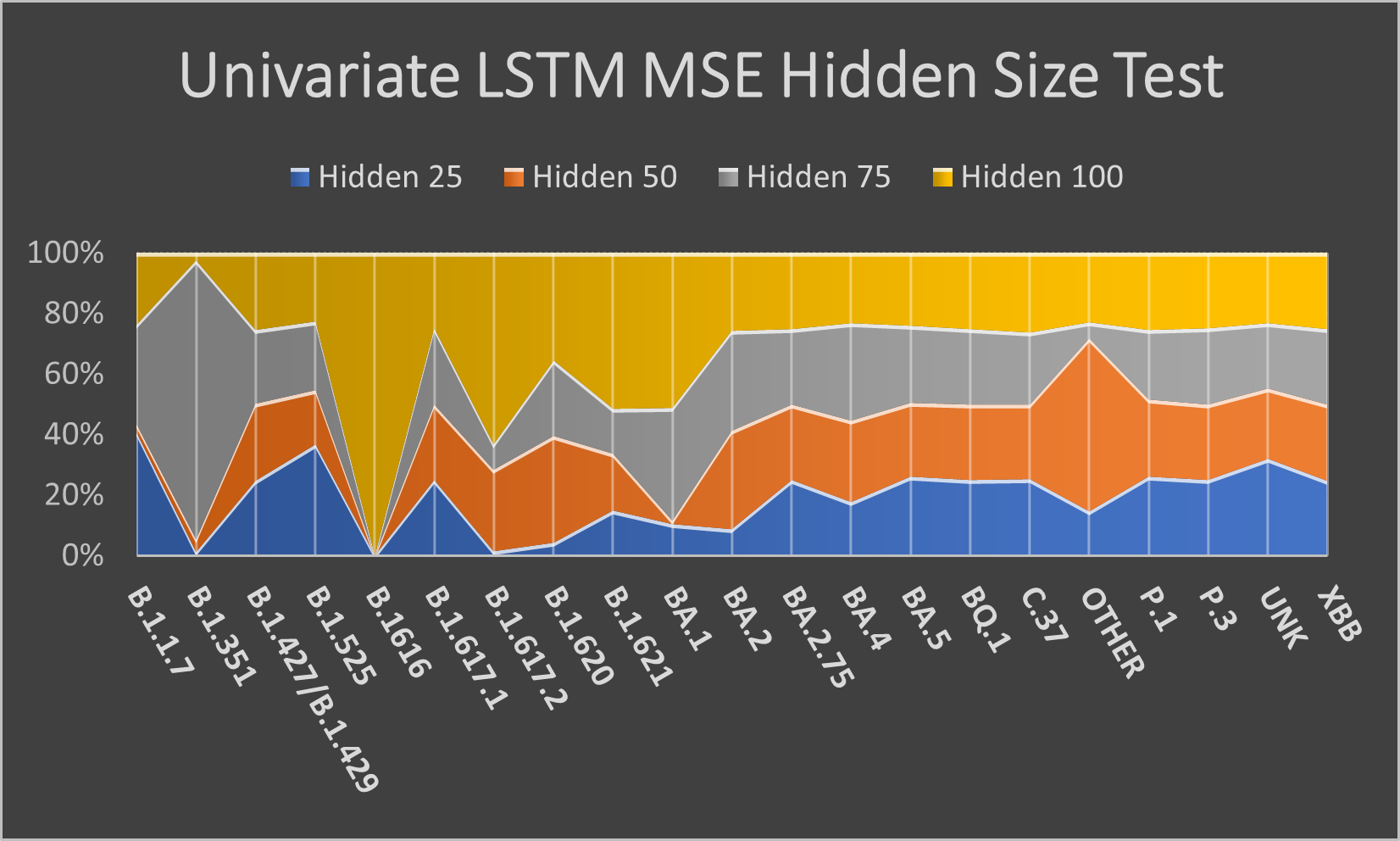}
	\caption{MSE value of Univariate Hidden Size Testing}
	\label{fig4}
\end{figure}

\begin{figure}[p]
	\centering	\includegraphics[width=0.9\textwidth]{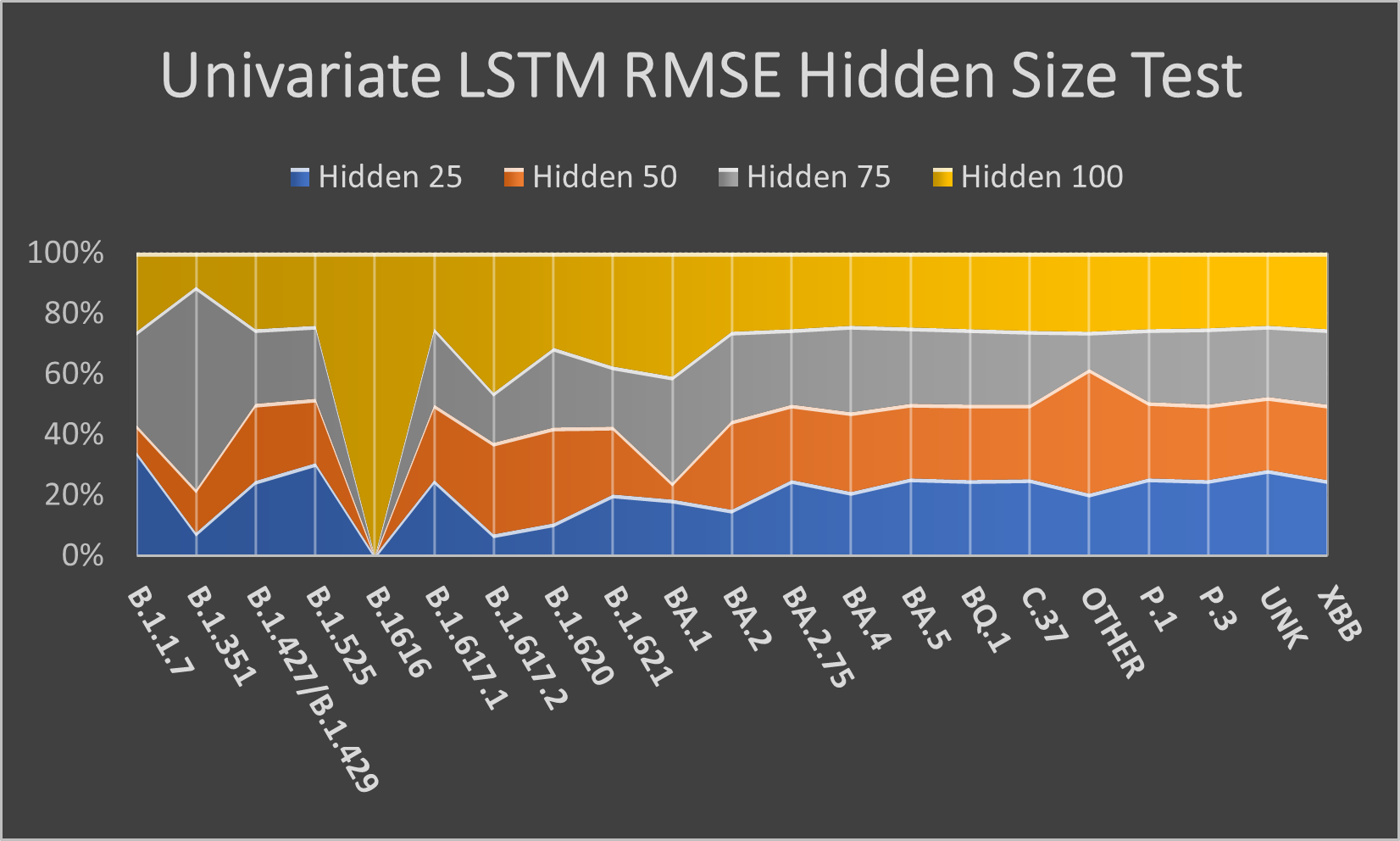}
	\caption{RMSE value of Univariate Hidden Size Testing}
	\label{fig5}
\end{figure}

Our model successfully predicted confirmed COVID-19 cases on each COVID-19 variant for one week ahead. On the Univariate hidden size test, our result indicates that most minimum loss values were obtained using hidden size 25, as shown in Figure \ref{fig4} and Figure \ref{fig5} for the LSTM model. The other models show the same result, with hidden size 25 giving the most minimum loss value from all variants compared to other hidden sizes.

\begin{figure}[p]
	\centering	\includegraphics[width=0.9\textwidth]{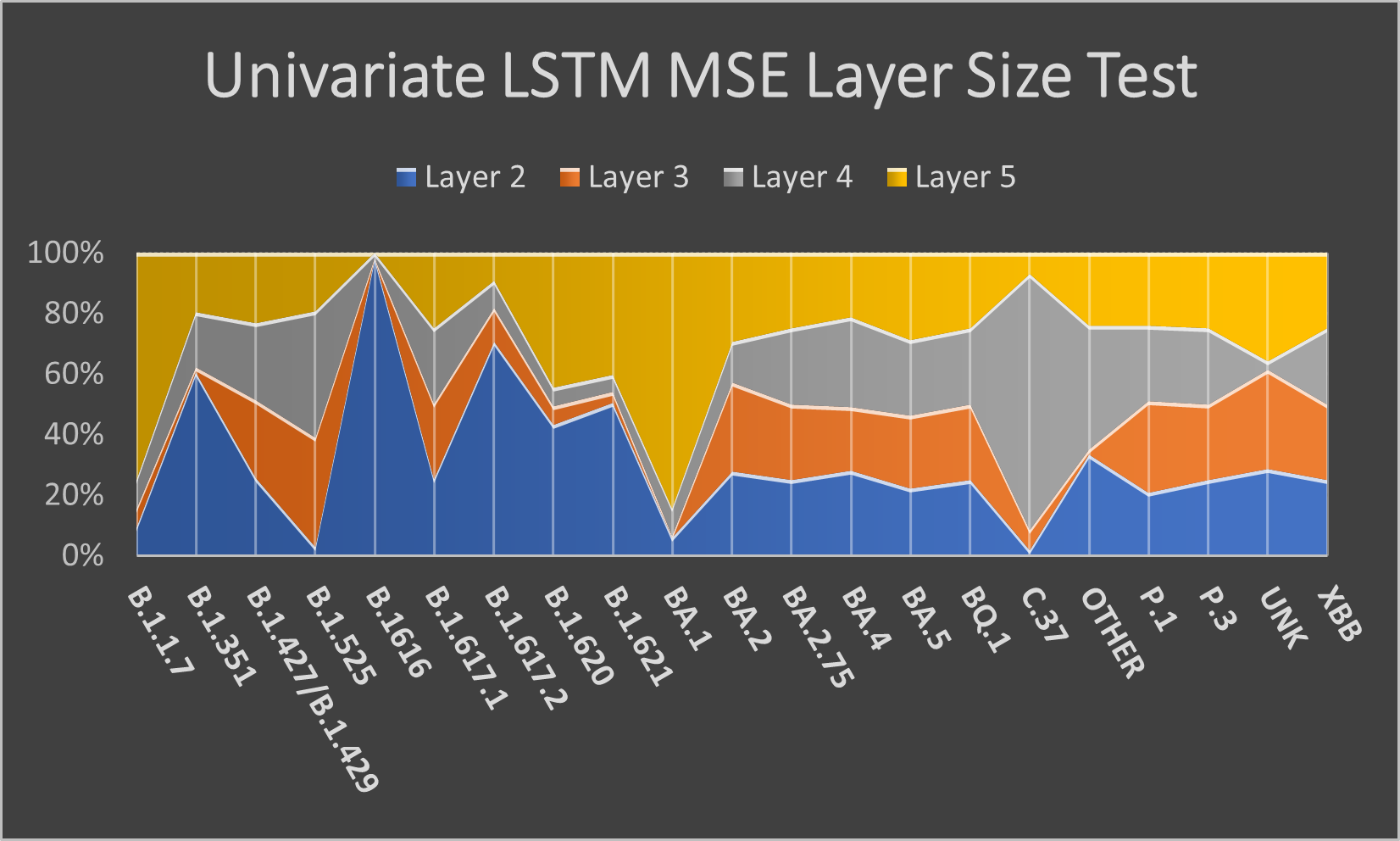}
	\caption{MSE value of Univariate Layer Size Testing}
	\label{fig6}
\end{figure}
\begin{figure}[p]
	\centering	\includegraphics[width=0.9\textwidth]{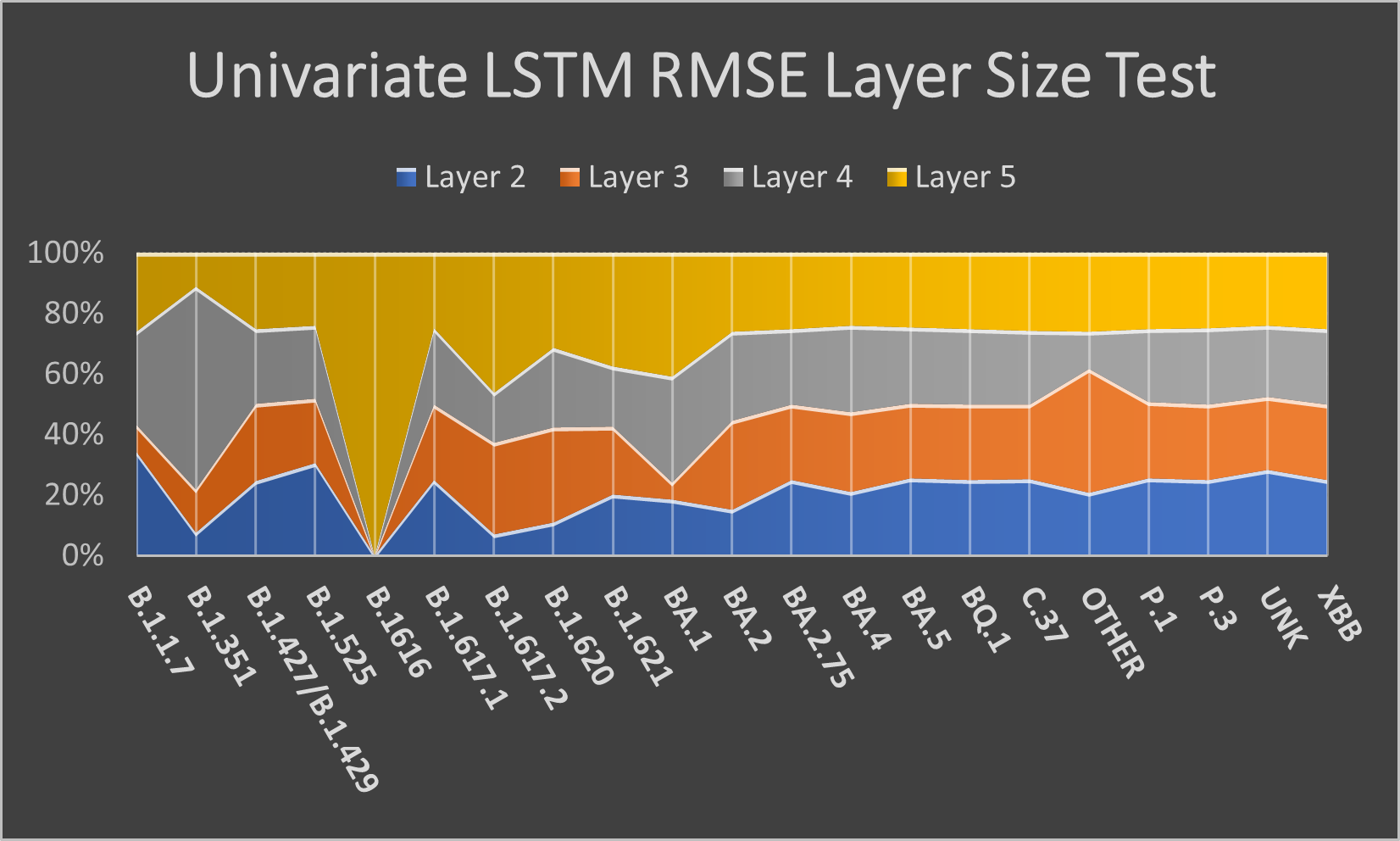}
	\caption{RMSE value of Univariate Layer Size Testing}
	\label{fig7}
\end{figure}

We use an area chart to represent the loss result, MSE, and RMSE, from all testing for each COVID-19 variant. Here the hidden size 25 provides a smaller area in Figure \ref{fig4} and Figure \ref{fig5} from all variants compared to other hidden size areas, meaning a smaller loss value is obtained mainly from using hidden size 25. There is an upward trend in the loss value if we look at the figure. The larger the hidden size used, the greater the loss value resulted.
\begin{table}[ht]
	\centering
	\caption{Minimum MSE Univariate Hidden Size Testing.}
	\label{table2}
	\begin{tabular}{@{}cccc@{}}
	\hline Variant& MSE LSTM& MSE BiLSTM& MSE RNN\\
    \hline
B.1.1.7& \textbf{0.062374894}& 0.15336436& 77.83555603\\
B.1.351& \textbf{6.31998E-05}& 8.71519E-05& 0.000206085\\
B.1.427/B.1.429& 9.12305E-05& 9.58116E-05& \textbf{8.00817E-05}\\
B.1.525& 0.000100649& \textbf{7.64034E-05}& 0.002185906\\
B.1616& \textbf{0}&	\textbf{0}&	\textbf{0}\\
B.1.617.1& 0.000100018&	9.44639E-05& \textbf{7.12258E-05}\\
B.1.617.2& \textbf{0.132738456}&	0.337521702&	23.38211823\\
B.1.620& \textbf{1.23469E-05}& 0.000103341& 0.000701719\\
B.1.621& 0.000109367& \textbf{7.1065E-05}&	0.000339199\\
BA.1& 2202.093994&	3290.208496& \textbf{2112.996582}\\
BA.2& \textbf{269726.9063}&	575534.8125& 42769.47266\\
BA.2.75& 443.3800964&	440.393219&	\textbf{400.5347595}\\
BA.4& 1436.005859& 1619.013306& \textbf{1277.752441}\\
BA.5& 721738.0625& 714504.375& \textbf{673358.12}\\
BQ.1& 17984.62109& 17956.88867&	\textbf{17573.03125}\\
C.37& 9.94712E-05&	\textbf{9.42242E-05}& 0.000351999\\
Other& 29.14533615&	\textbf{3.766394138}& 80.57791901\\
P.1& 0.002713137& \textbf{0.001758261}& 0.003104303\\
P.3& 0.000377027& 0.00037388&	\textbf{0.000313186}\\
UNK& 0.024996253& \textbf{0.023208618}&	0.084605187\\
XBB& \textbf{555.9832153}& 561.4128418&	561.8463135\\
    \hline
	\end{tabular}
\end{table}

From all hidden size tests, 25 to 100, we take the minimum loss value obtained for each COVID-19 variant. The result is shown in Table \ref{table2} and Table \ref{table3}. The bolded values on the tables indicate that these are the minimum values of the three models tested for each COVID-19 variant. When we compare all models, RNN gave the most minimum loss values than the other two from all variants tested.  
\begin{table}[ht]
	\centering
	\caption{Minimum RMSE Univariate Hidden Size Testing.}
	\label{table3}
	\begin{tabular}{@{}cccc@{}}
	\hline Variant& RMSE LSTM& RMSE BiLSTM& RMSE RNN \\
    \hline
B.1.1.7& \textbf{0.24974966}& 0.391617626& 8.822445869\\
B.1.351& \textbf{0.007949829}& 0.009335518& 0.014355659\\
B.1.427/B.1.429& 0.009551466& 0.009788342& \textbf{0.00894884}\\
B.1.525& 0.010032415& \textbf{0.008740902}& 0.046753675\\
B.1616& \textbf{0}&	\textbf{0}&	\textbf{0}\\
B.1.617.1& 0.010000885&	0.009719253& \textbf{0.00843954}\\
B.1.617.2& \textbf{0.364332885}&	0.580966175& 4.835505962\\
B.1.620& 0.003513813& \textbf{0.010165691}& 0.026489971\\
B.1.621& 0.010457873&	0.008430004&	0.018417358\\
BA.1& 46.92647552& 57.36034012& \textbf{45.96734238}\\
BA.2& 519.352417& 758.6401367& \textbf{206.8078156}\\
BA.2.75& 21.05659294& 20.98554802& \textbf{20.01336479}\\
BA.4& 37.89466858& 40.23696518&	\textbf{35.74566269}\\
BA.5& 849.5516968&	845.2836304& \textbf{820.5839844}\\
BQ.1& 134.1067505&	134.0033112& \textbf{132.5633087}\\
C.37& 0.009973525&	0.009706914& \textbf{0.018761635}\\
Other& 5.398642063&	\textbf{1.940719962}& 8.976520538\\
P.1& 0.052087784& \textbf{0.041931629}& 0.055716276\\
P.3& 0.019417183& 0.019335989& \textbf{0.017697062}\\
UNK& 0.158102036& \textbf{0.15234375}& 0.290869713\\
XBB& \textbf{23.57929611}& 23.69415283& 23.70329666\\
    \hline
	\end{tabular}
\end{table}

As we can see from both Table \ref{table2} and Table \ref{table3}, for variant B.1616, the loss value is 0, meaning the three models predict the cases accurately. This perfect score happened because most of its data contained 0 cases and no higher than 1 case.

Unlike the hidden size test, adding more layers resulting in LSTM gave the most minimum loss value obtained here on the layer size test. Using the configuration of hidden size 25 and layer size 3, LSTM topped the other two models. Figure \ref{fig6} and Figure \ref{fig7} show different results of LSTM using layer sizes 2 to 5. The same result also goes for BiLSTM and RNN.

From the figure, the overall area of layer size 3 is smaller than other layer sizes, which means that using layer size three results in a smaller or minimum loss value. Increasing the number of layers does not help reduce the loss value. Instead, it increases its value, as shown from the area of layer size 5. BiLSTM and RNN also gave the same result as LSTM, with layer size three as the optimal configuration.

\begin{figure}[p]
	\centering	\includegraphics[width=1.0\textwidth]{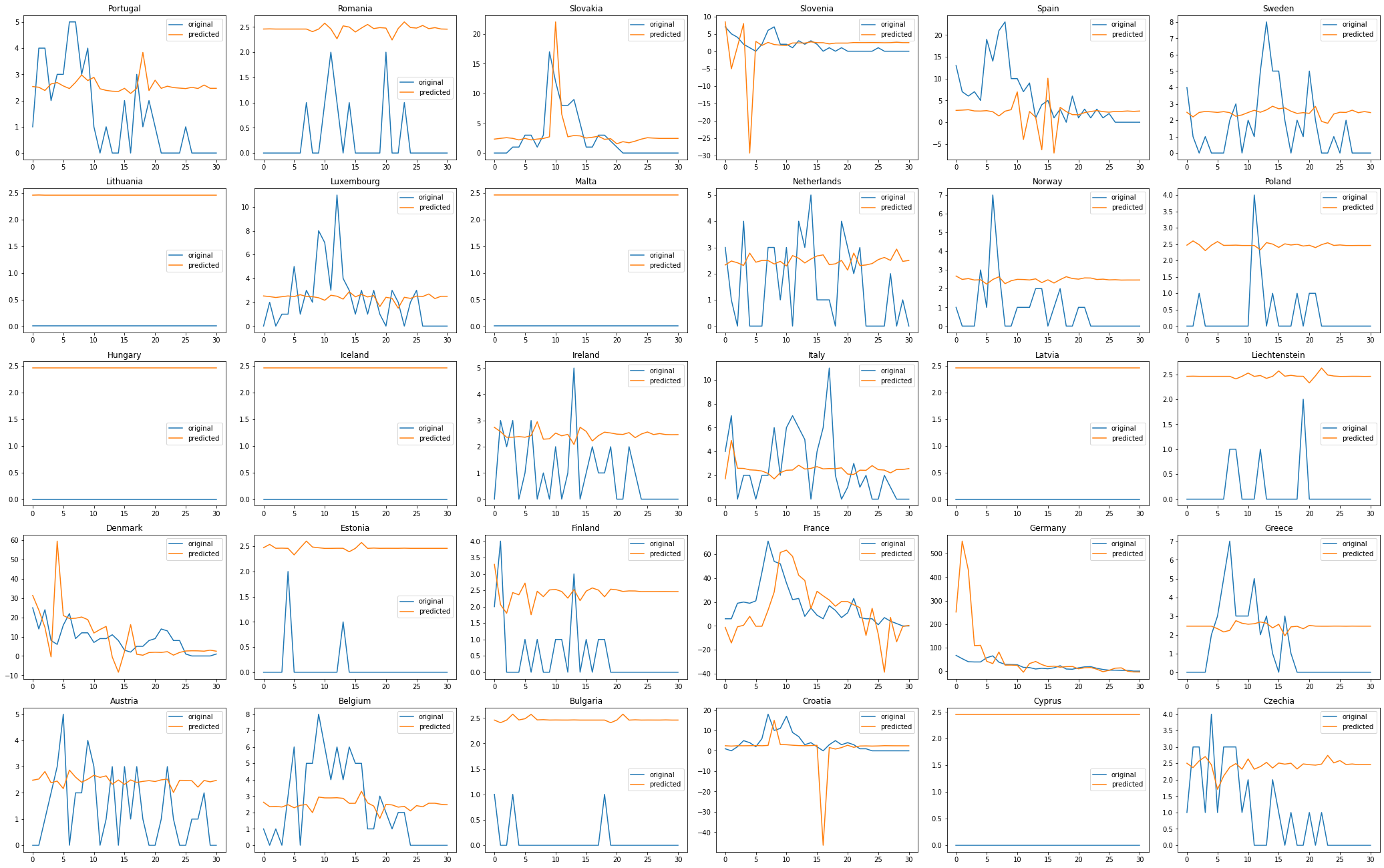}
	\caption{Univariate LSTM using hidden 25 layer 3 on "Other" variant}
	\label{fig8}
\end{figure}

\begin{figure}[p]
	\centering	\includegraphics[width=1.0\textwidth]{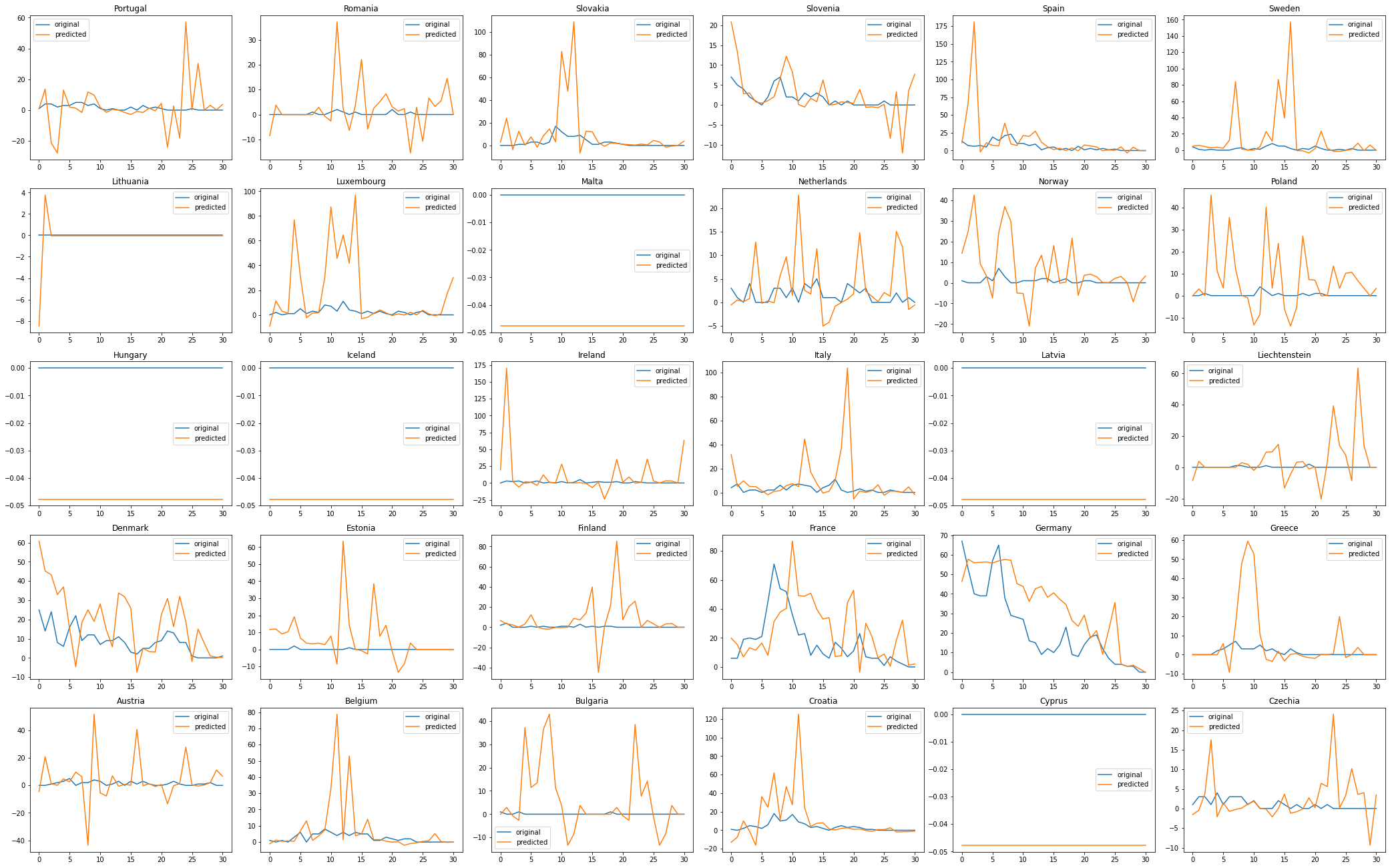}
	\caption{Univariate BiLSTM using hidden 25 layer 3 on "Other" variant}
	\label{fig9}
\end{figure}

\begin{figure}[ht]
	\centering	\includegraphics[width=1.0\textwidth]{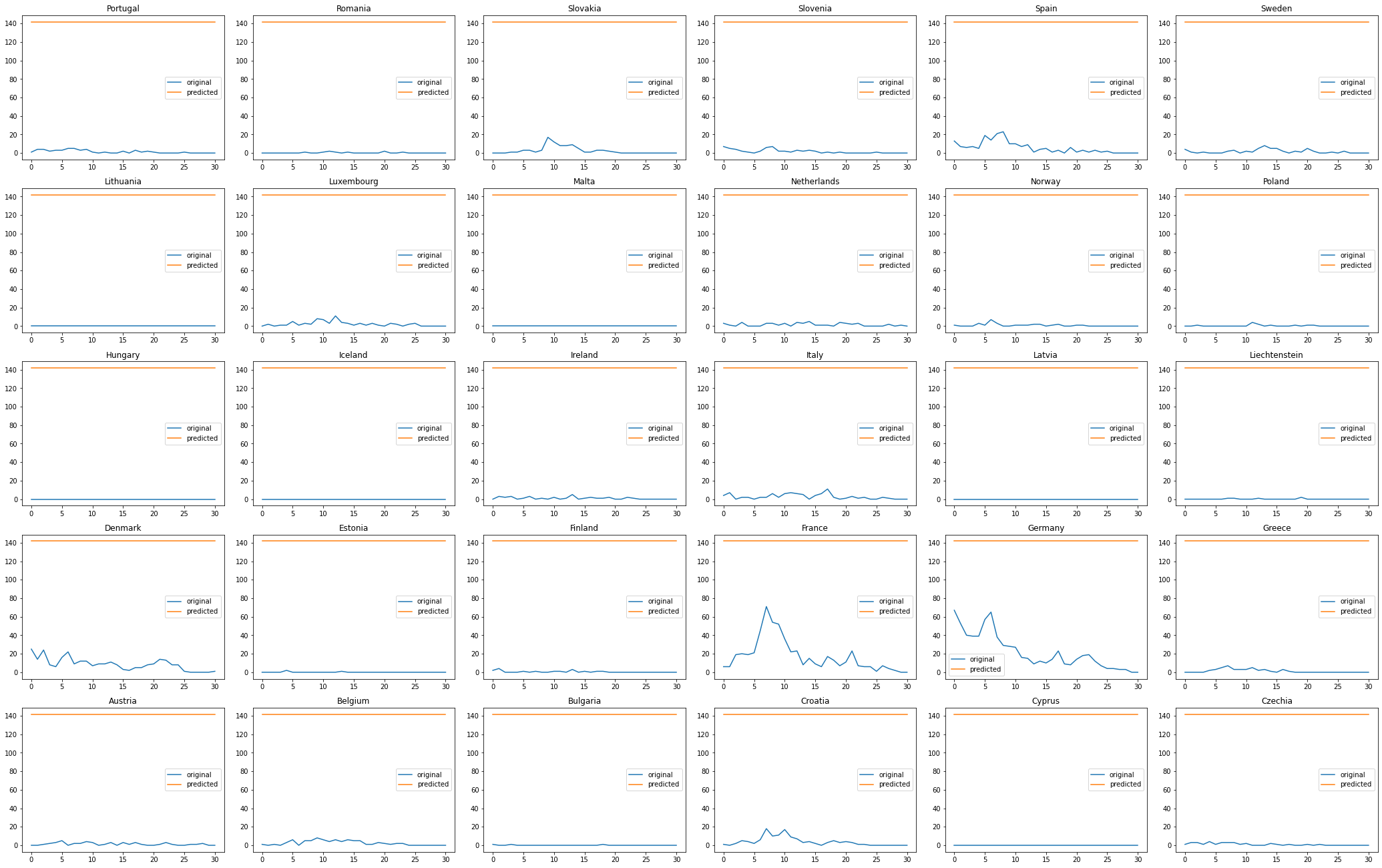}
	\caption{Univariate RNN using hidden 25 layer 3 on "Other" variant}
	\label{fig10}
\end{figure}

Our prediction chart contains 30 countries in Europe corresponding to the dataset we used. The country started from Austria at the bottom left of the last row, then continued to the right and started again from the left, repeated until the upper rows. We show the prediction chart from our optimal configuration, using hidden size 25 and layer size 3, divided into three figures based on the model implemented, Figure \ref{fig8} for LSTM, Figure \ref{fig9} for BiLSTM and Figure \ref{fig10} for RNN. All of them are predictions for the "Other" variant only. LSTM gave the closest prediction to the real data compared to the other two models. On the other hand, BiLSTM and RNN predict way too far from their original data, resulting in a larger loss value than LSTM. 

Table \ref{table4} and \ref{table5} contain the minimum loss value of the overall Univariate layer size test. Most loss values on the tables come from the LSTM model, followed by BiLSTM and RNN. These results further proved that RNN is unsuitable for long-term patterns, which is the fundamental reason LSTM developed to overcome this issue \cite{Li2018}. RNN may work well on single-layer architecture, but multilayered RNNs still cannot beat LSTM on the same configuration. Thus we concluded on the Univariate test that the optimal configuration is using the LSTM model with hidden size 25 and layer size 3. 
\begin{table}[ht]
	\centering
	\caption{Minimum Univariate MSE Layer Size Testing.}
	\label{table4}
	\begin{tabular}{@{}cccc@{}}
	\hline Variant& MSE LSTM& MSE BiLSTM& MSE RNN \\
    \hline
B.1.1.7& \textbf{0.285582066}& 0.048046913&	0.697912753\\
B.1.351& \textbf{0.000278501}& 0.000596882& 0.050010335\\
B.1.427/B.1.429& \textbf{8.72087E-05}& 9.60345E-05&	0.000313051\\
B.1.525& \textbf{9.04E-06}&	0.000110109& 0.000420311\\
B.1616& \textbf{0}&	\textbf{0}&	\textbf{0}\\
B.1.617.1& 9.93619E-05&	\textbf{9.77854E-05}& 0.000324554\\
B.1.617.2& 0.272779733&	\textbf{0.12485259}&	37.21570206\\
B.1.620& 0.000102891& \textbf{9.17507E-05}& 0.000704678\\
B.1.621& \textbf{2.74123E-05}& 0.000104615&	0.001252985\\
BA.1& 51.86838531& \textbf{21.66649628}&	135.2298737\\
BA.2& \textbf{1069615.5}& 1938490.125&	604102.8125\\
BA.2.75& 445.1516724& 445.1611938& \textbf{431.5044556}\\
BA.4& 2279.553711&	1890.885254& \textbf{1747.677124}\\
BA.5& 680154.625&	693546.5625& \textbf{643830.4375}\\
BQ.1& \textbf{18005.15234}&	18011.95898& 18012.54688\\
C.37& \textbf{2.77505E-05}&	8.96806E-05& 0.001253018\\
Other& \textbf{4.598902225}& 275.0593262& 1948.237183\\
P.1& 0.002071609& 0.002713932&	\textbf{0.013807666}\\
P.3& 0.000377038& 0.000377015& \textbf{0.000313259}\\
UNK& \textbf{0.003363617}&	0.01490442&	0.085280031\\
XBB& 561.8049927& 561.9882813& \textbf{557.6511841}\\
    \hline
	\end{tabular}
\end{table}
\begin{table}[ht]
	\centering
	\caption{Minimum Univariate RMSE Layer Size Testing.}
	\label{table5}
	\begin{tabular}{@{}cccc@{}}
	\hline Variant& RMSE LSTM& RMSE BiLSTM& RMSE RNN \\
    \hline
B.1.1.7& 0.534398794& \textbf{0.219196066}& 0.835411727\\
B.1.351& \textbf{0.016688347}& 0.024431167& 0.223629907\\
B.1.427/B.1.429& \textbf{0.009338558}& 0.009799719& 0.01769325\\
B.1.525& \textbf{0.003005981}& 0.010493278& 0.020501493\\
B.1616&	\textbf{0}&	\textbf{0}&	\textbf{0}\\
B.1.617.1& 0.009968042&	\textbf{0.009888649}& 0.018015385\\
B.1.617.2& 0.522283196&	\textbf{0.353344858}& 6.100467205\\
B.1.620& 0.010143518& \textbf{0.009578658}& 0.026545763\\
B.1.621& \textbf{0.005235672}& 0.010228172& 0.03539753\\
BA.1& 7.201971054& \textbf{4.654728413}& 11.62883759\\
BA.2& 1034.222168&	1392.296753& \textbf{777.2405396}\\
BA.2.75& 21.09861755& 21.09884262& \textbf{20.772686}\\
BA.4& 47.74467087& 43.48431015& \textbf{41.80522919}\\
BA.5& 824.7149048&	832.7944336& \textbf{802.3904419}\\
BQ.1& \textbf{134.1832733}&	134.2086334& 134.2108307\\
C.37& 0.005267873&	0.009469985& \textbf{0.035397992}\\
Other& \textbf{2.144505024}& 16.58491325& 44.13883972\\
P.1& \textbf{0.045514934}& 0.052095413& 0.11750602\\
P.3& 0.019417465& 0.019416876& \textbf{0.017699111}\\
UNK& \textbf{0.057996694}& 0.122083656& 0.292027444\\
XBB& 23.702425&	23.7062912&	\textbf{23.61463928}\\
    \hline
	\end{tabular}
\end{table}

By comparing the loss value of the hidden size test with the layer size test, most loss values for each variant decrease in the layer size test. Although not all variants, for example, the B.1.427/B.1.429 variant has increased its loss value. However, the B.1616 variant still provides a 0 loss value even in the layer size test. 

\subsection{Multivariate Test}
\begin{figure}[p]
	\centering	\includegraphics[width=0.9\textwidth]{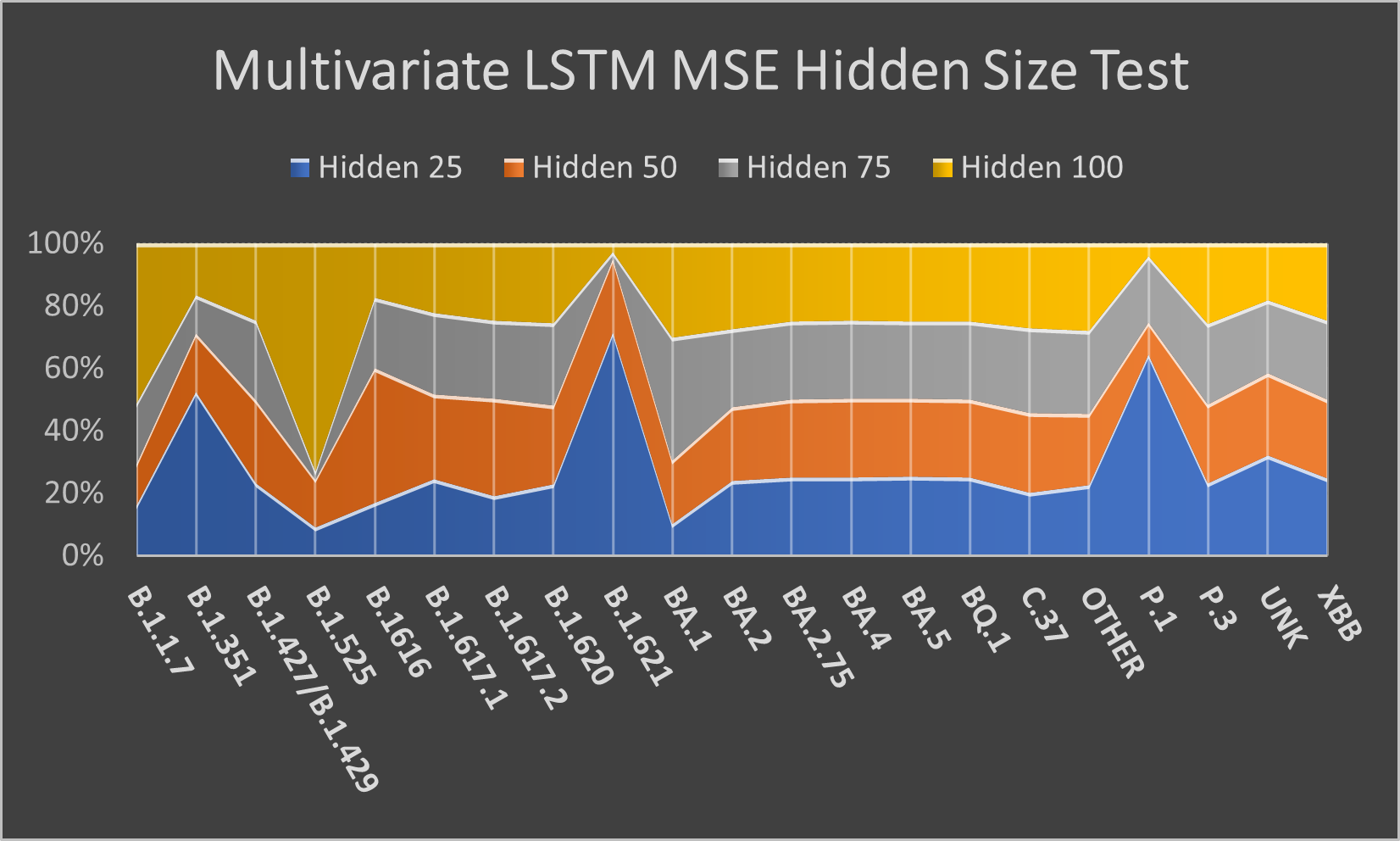}
	\caption{MSE value of Multivariate Hidden Size Testing}
	\label{fig11}
\end{figure}

\begin{figure}[p]
	\centering	\includegraphics[width=0.9\textwidth]{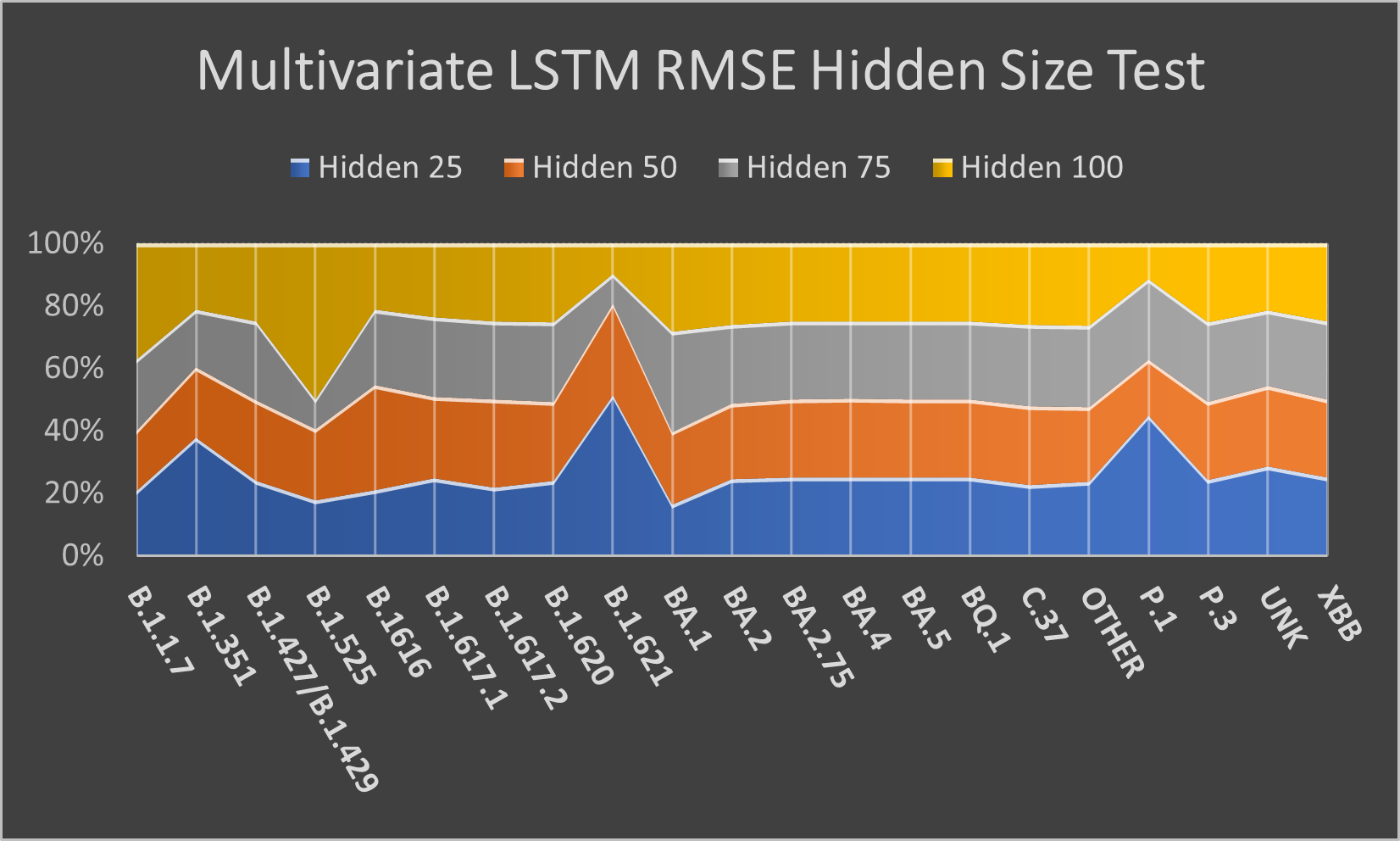}
	\caption{RMSE value of Multivariate Hidden Size Testing}
	\label{fig12}
\end{figure}

For the Multivariate hidden size test, our result is similar to the Univariate, with hidden size 25 giving the most minimum loss value obtained from all variants. Here in Figure \ref{fig11} and Figure \ref{fig12} for LSTM, we can see that from all COVID-19 variants, most minimum loss values were obtained by using hidden size 25, proved by its smaller area on the chart. Interestingly though, among all the tested hidden sizes, they all have a slight difference in the frequency of minimum loss values obtained between each other. Nevertheless, it resulted in a quite balanced distribution on the area charts. The upward trend remains in the Multivariate hidden size test, just like the Univariate. The difference is not too much. Thus, it is not too noticeable. 

\begin{figure}[p]
	\centering	\includegraphics[width=0.9\textwidth]{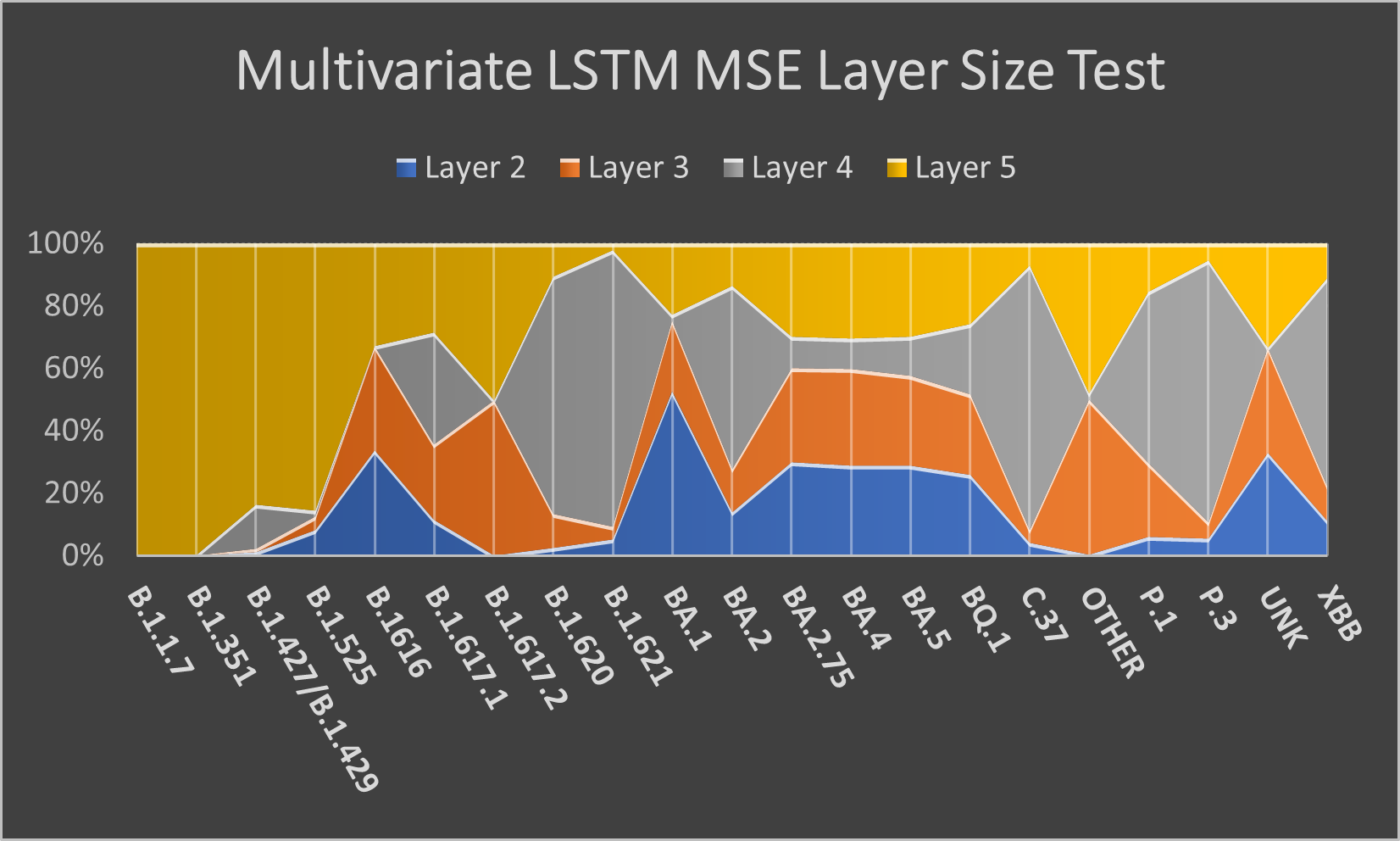}
	\caption{MSE value of Multivariate Layer Size Testing}
	\label{fig13}
\end{figure}
\begin{figure}[p]
	\centering	\includegraphics[width=0.9\textwidth]{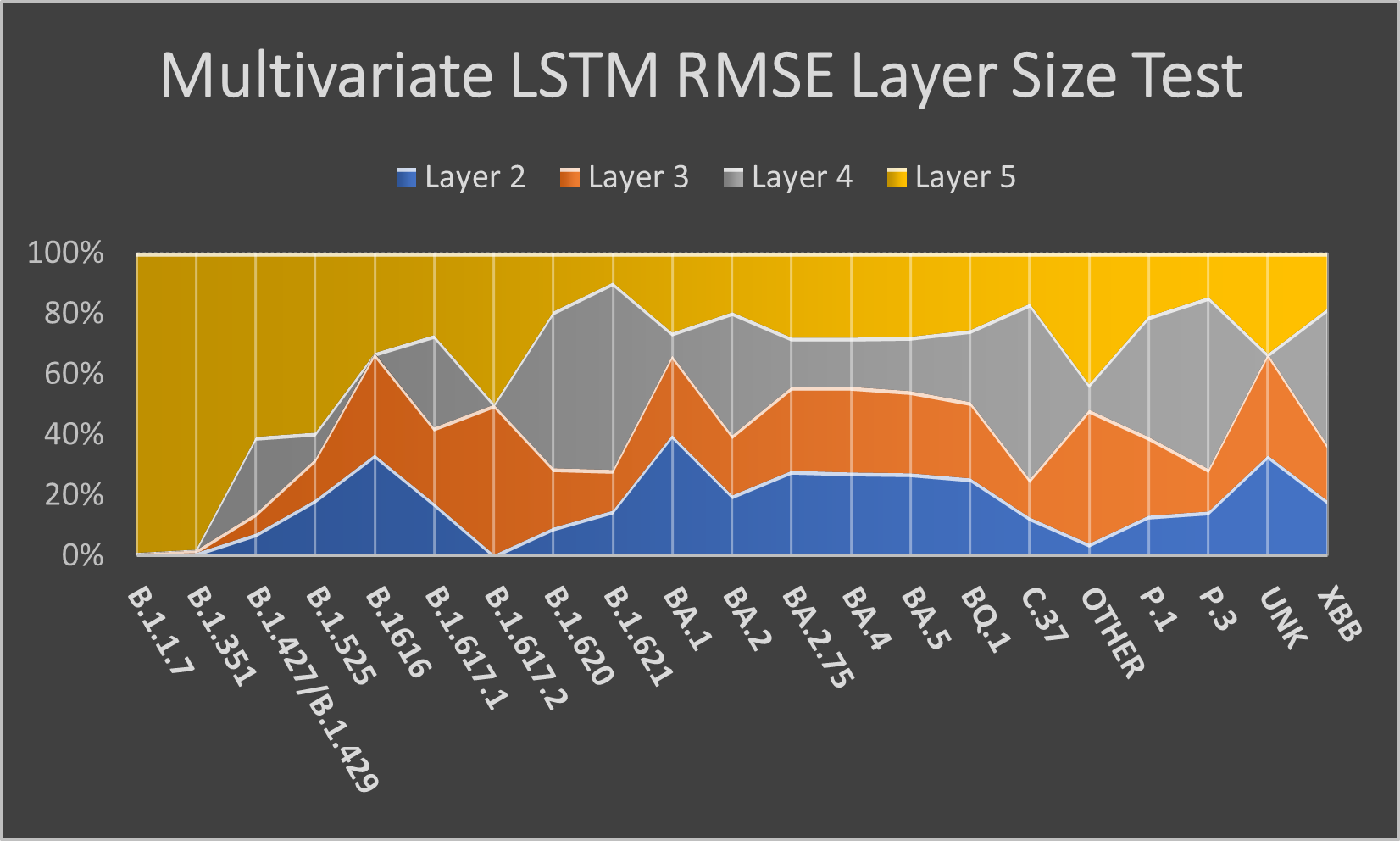}
	\caption{RMSE value of Multivariate Layer Size Testing}
	\label{fig14}
\end{figure}

The minimum loss value from all hidden size tests is shown in Table \ref{table6} and Table \ref{table7}. Judging by looking at the bold value on the tables, here on Multivariate, the LSTM performs better. Different from Univariate cases where RNN outperforms LSTM. The frequency of the minimum loss value obtained by LSTM for each COVID-19 variant exceeds the RNN. Another example is that RNN works well only on Univariate single-layer cases. 
\begin{table}[ht]
	\centering
	\caption{Minimum Multivariate MSE Hidden Size Testing.}
	\label{table6}
	\begin{tabular}{@{}cccc@{}}
	\hline Variant& MSE LSTM& MSE BiLSTM& MSE RNN \\
    \hline
B.1.1.7& \textbf{0.04227148}& 0.060507733&	3.309589386\\ B.1.351& 0.002829222& \textbf{0.002783381}&	0.003932106\\  B.1.427/B.1.429& \textbf{8.19067E-06}& 9.41505E-06& 8.33682E-06\\
B.1.525& \textbf{2.63134E-05}& 2.73274E-05& 0.000232415\\
B.1616& 4.02546E-06& \textbf{3.81788E-06}& 4.01232E-06\\
B.1.617.1& 3.59397E-05&	\textbf{2.58086E-05}& 2.88424E-05\\
B.1.617.2& 51.76334& 9.645352364& \textbf{9.133572578}\\
B.1.620& \textbf{2.18086E-05}& 2.54787E-05& 4.9359E-05\\
B.1.621& 8.94553E-06& \textbf{1.20327E-05}& 7.92159E-05\\
BA.1& 6295.939941& 11615.52539& \textbf{4038.050537}\\
BA.2& \textbf{371409.6875}& 375215.8438& 373718.8438\\
BA.2.75& 1332.728638&	1333.700928& \textbf{1330.381226}\\
BA.4& \textbf{6960.57959}& 7049.269043&	7252.370117\\
BA.5& \textbf{1516468}& 1522720.75&	1532071.125\\
BQ.1& 20797.63477&	20828.27734& \textbf{20750.55469}\\
C.37& \textbf{2.18528E-05}&	2.7185E-05&	2.97164E-05\\
Other& \textbf{67.05672455}& 74.06816101& 88.95769501\\
P.1& \textbf{0.00020777}& 0.000266735& 0.000521\\
P.3& 1.87686E-05& 2.0501E-05&	\textbf{1.81369E-05}\\
UNK& \textbf{80732.13281}& 83808.39844&	81823.41406\\
XBB& 93.5736084& 93.7795105& \textbf{92.58743286}\\
    \hline
	\end{tabular}
\end{table}
\begin{table}[ht]
	\centering
	\caption{Minimum Multivariate RMSE Hidden Size Testing.}
	\label{table7}
	\begin{tabular}{@{}cccc@{}}
	\hline Variant& RMSE LSTM& RMSE BiLSTM& RMSE RNN \\
    \hline
B.1.1.7& \textbf{0.205600291}& 0.245983198& 1.819227695\\
B.1.351& 0.053190429& \textbf{0.052757762}& 0.062706508\\
B.1.427/B.1.429& \textbf{0.002861935}& 0.003068395&	0.002887355\\
B.1.525& \textbf{0.00512966}& 0.005227563&	0.015245159\\
B.1616& 0.002006355& \textbf{0.001953939}&	0.002003078\\
B.1.617.1& 0.005994972&	\textbf{0.005080214}& 0.005370509\\
B.1.617.2& 7.194674492&	3.105696678& \textbf{3.02218008}\\
B.1.620& \textbf{0.004669973}& 0.005047639& 0.007025594\\
B.1.621& \textbf{0.002990909}& 0.003468824&	0.008900332\\
BA.1& 79.34696198&	107.7753448& \textbf{63.54565811}\\
BA.2& \textbf{609.4338989}&	612.548645&	611.3255005\\
BA.2.75& 36.50655746&	36.51987076& \textbf{36.47439194}\\
BA.4& \textbf{83.43008423}&	83.95992279&	85.16085052\\
BA.5& \textbf{1231.449585}&	1233.985718&	1237.768555\\
BQ.1& 144.2138519&	144.3200531& \textbf{144.0505219}\\
C.37& \textbf{0.004674699}&	0.005213923& 0.005451273\\
Other& \textbf{8.188817024}& 8.606286049& 9.431738853\\
P.1& \textbf{0.014414218}& 0.016332025& 0.022845136\\
P.3& 0.004332275&	0.004527803& \textbf{0.004258746}\\
UNK& \textbf{284.1340027}& 289.4967957&	286.0479126\\
XBB& 9.673345566&	9.683981895& \textbf{9.622236252}\\
    \hline
	\end{tabular}
\end{table}

Another interesting thing to observe is that variant B.1616 did not give a 0 loss value like the previous result on the Univariate test. The three models failed to predict accurately. Our assumption with the simultaneous inclusion of varied data into the model caused this result, unlike Univariate, which enters the data one by one, allowing for a better training process. However, the overall result proves to be better than the Univariate one. 

Next, on the layer size test, more layers added to the model did decrease the loss value. However, our optimal result was on a smaller layer size. Instead, it is on layer size 4. This result is represented in Figure \ref{fig13} and Figure \ref{fig14} for LSTM. Using a layer size of 5 makes it the worst configuration, with the many COVID-19 variant resulting in the highest loss value compared to the others. The other two models brought the same result, with layer size four as the optimal configuration.

\begin{figure}[p]
	\centering	\includegraphics[width=1.0\textwidth]{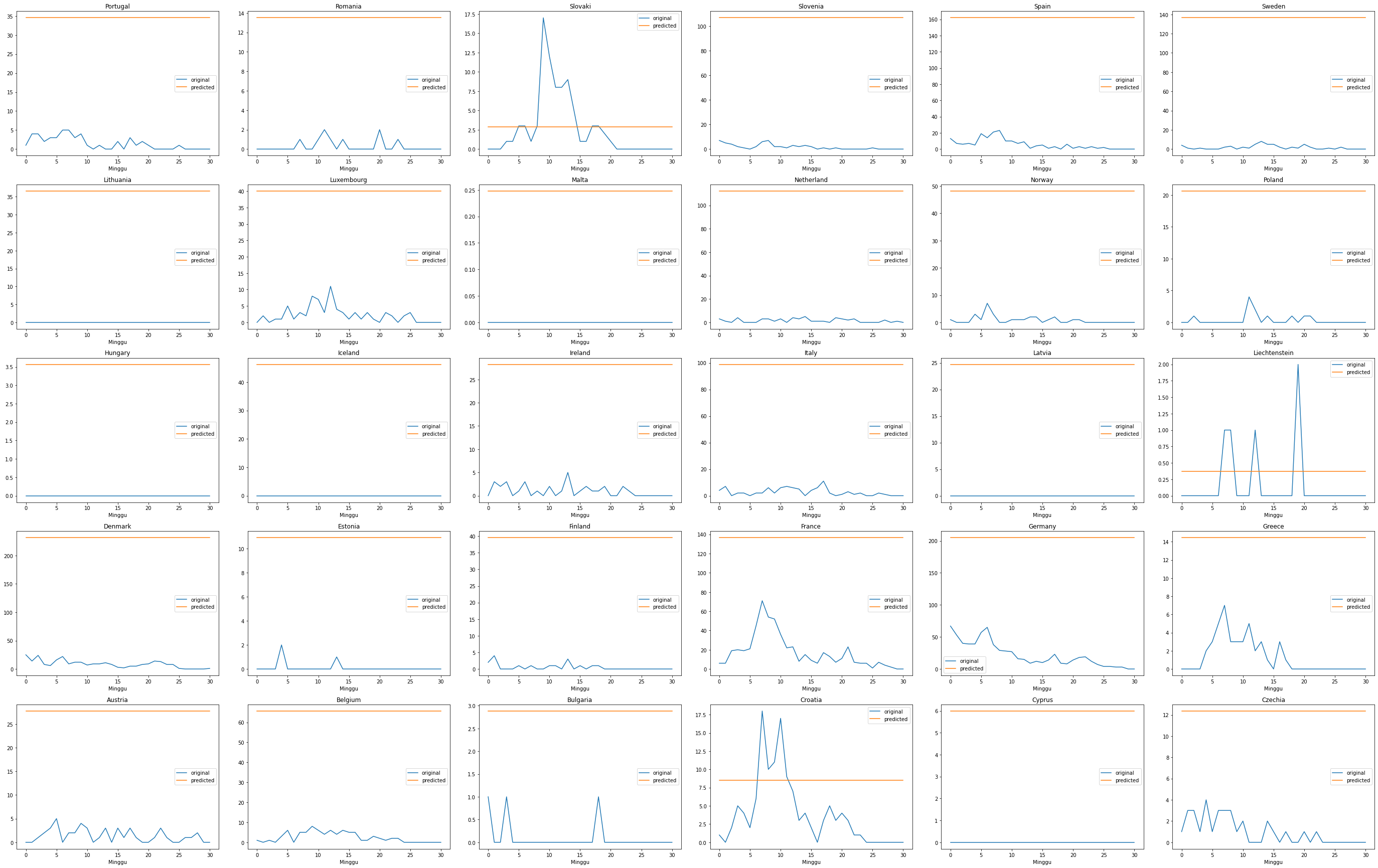}
	\caption{Multivariate LSTM using hidden 25 layer 4 on "Other" variant}
	\label{fig15}
\end{figure}

\begin{figure}[p]
	\centering	\includegraphics[width=1.0\textwidth]{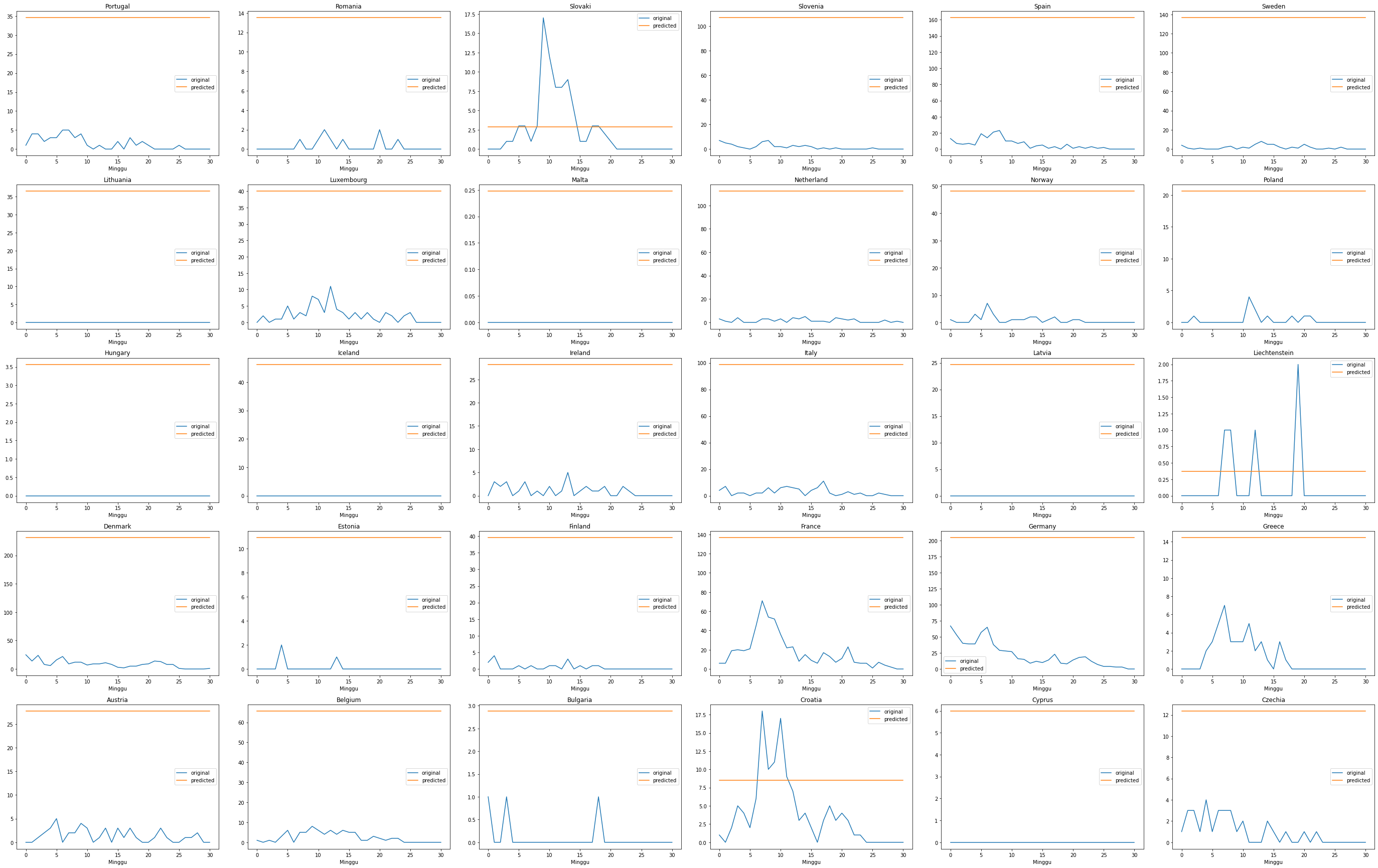}
	\caption{Multivariate BiLSTM using hidden 25 layer 4 on "Other" variant}
	\label{fig16}
\end{figure}

\begin{figure}[ht]
	\centering	\includegraphics[width=1.0\textwidth]{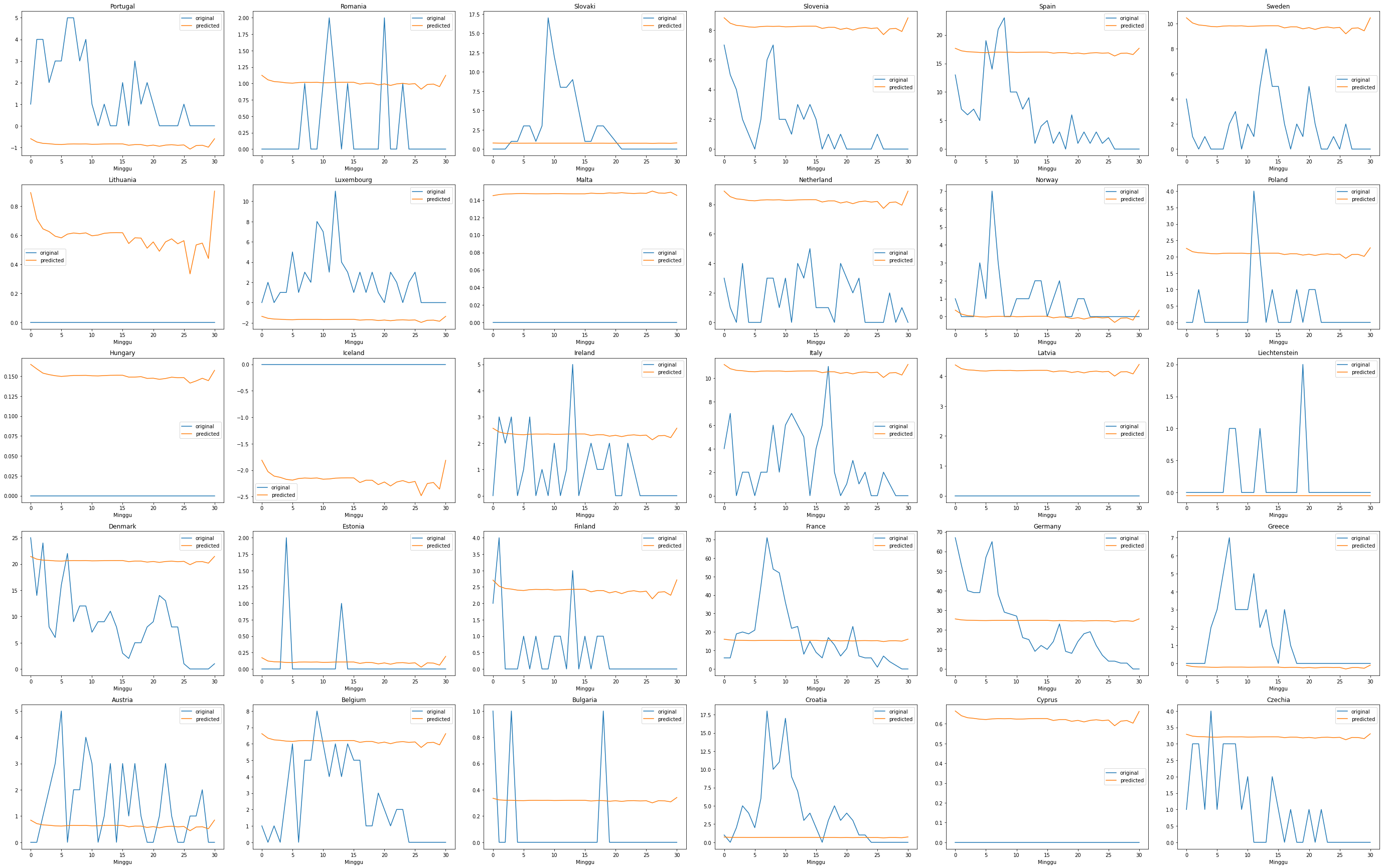}
	\caption{Multivariate RNN using hidden 25 layer 4 on "Other" variant}
	\label{fig17}
\end{figure}

Identical to Univariate, here we show our prediction chart for each model shown in Figure \ref{fig15} for LSTM, Figure \ref{fig16} for BiLSTM, and Figure \ref{fig17} for RNN using our optimal configuration for Multivariate, which is hidden size 25 and layer size 4. All of them are "Other" variant predictions. The prediction results in this test are rather flatter, different from Univariate, which still shows a tendency to follow the shape of the data. However, the predicted value is larger than the original data, unlike in Univariate, which tends to be smaller than the original data.


Our prediction result for all variants, both Univariate and Multivariate tests, gave unfavourable results, with most predicting far beyond or even below the original data. From this problem, we acknowledge these results due to the dataset containing many zero values, thus influencing our prediction result. TThis problem can be solved by increasing the amount of data used or trying other normalization methods.

The minimum loss value for each variant from the entire Multivariate layer size test is shown in Table \ref{table8} and Table \ref{table9}. The loss value in our Multivariate layer size test provides a loss value with an upward trend compared to the hidden size test. Although this loss value is already the minimum value of the entire layer size test, this result is inversely proportional to the Univariate test, which has a downward trend with the addition of layer size. 

When we calculate the frequency of minimum loss value for each COVID-19 variant, we found that both LSTM and BiLSTM have the same frequency of minimum loss value. Meaning that both LSTM and BiLSTM perform approximately equally well. In reality, BiLSTM should outperform the LSTM and provides better result and performance \cite{Siami2019}; in our case, the frequency of minimum loss value should be larger. 
\begin{table}[ht]
	\centering
	\caption{Minimum Multivariate MSE Layer Size Testing.}
	\label{table8}
	\begin{tabular}{@{}cccc@{}}
	\hline Variant& MSE LSTM& MSE BiLSTM& MSE RNN \\
    \hline
B.1.1.7& \textbf{0.030201122}& 0.087667666&	0.125225887\\
B.1.351& \textbf{0.000407222}& 0.004255008& 0.006206059\\
B.1.427/B.1.429&  7.28716E-06& \textbf{6.82E-06}& 9.39745E-06\\
B.1.525& \textbf{4.07541E-05}& 0.000141382& 0.001054534\\
B.1616&	\textbf{0}&	\textbf{0}&	\textbf{0}\\
B.1.617.1& 3.01427E-05&	2.76918E-05& \textbf{2.18035E-05}\\
B.1.617.2& 0.702166438& \textbf{0.168214917}& 296.4614868\\
B.1.620& 2.63529E-05& \textbf{2.30556E-05}& 2.60507E-05\\
B.1.621& \textbf{1.16705E-05}& 5.12911E-05&	5.84191E-05\\
BA.1& 1181.40564& \textbf{601.9147949}&	13268.86914\\
BA.2& \textbf{374334.625}&	374705.1563&	374586.1563\\
BA.2.75& 444.7804871&	445.1540833& \textbf{431.7670898}\\
BA.4& \textbf{2412.172363}&	3070.185791& 4485.07959\\
BA.5& \textbf{650916.25}& 726488.125& 813796.4375\\
BQ.1& 18016.07813& \textbf{18001.36328}& 18012.54688\\
C.37& \textbf{2.55481E-05}&	2.98774E-05&	3.05288E-05\\
Other& 	54.60609055& \textbf{28.08243752}&	46.32154465\\
P.1& \textbf{0.000335828}&	0.000404327&	0.025433218\\
P.3& 1.98145E-05& \textbf{1.66194E-05}&	1.98912E-05\\
UNK& 0.082096986& \textbf{0.000518148}&	0.08528506\\
XBB& 93.92437744& \textbf{93.88194275}&	93.96388245\\
    \hline
	\end{tabular}
\end{table}
\begin{table}[ht]
	\centering
	\caption{Minimum Multivariate Layer Size RMSE Testing.}
	\label{table9}
	\begin{tabular}{@{}cccc@{}}
	\hline Variant& RMSE LSTM& RMSE BiLSTM& RMSE RNN \\
    \hline
B.1.1.7& \textbf{0.173784703}& 0.296087265&	0.353872687\\
B.1.351& \textbf{0.020179749}& 0.065230429&	0.07877855\\
B.1.427/B.1.429& 0.002699475& \textbf{0.002611829}& 0.003065527\\
B.1.525& \textbf{0.006383896}& 0.011890434&	0.032473594\\
B.1616&	\textbf{0}& \textbf{0}&	\textbf{0}\\
B.1.617.1& 0.005490233&	0.005262296& \textbf{0.004669424}\\
B.1.617.2& 0.837953746&	\textbf{0.410140127}& 17.21805763\\
B.1.620&0.005133511& \textbf{0.004801628}&	0.005103985\\
B.1.621&\textbf{0.003416212}& 0.007161778&	0.007643239\\
BA.1& 34.37158203& \textbf{24.53395271}&	115.1905746\\
BA.2& \textbf{611.8289185}&	612.1316528&	612.0344238\\
BA.2.75& 21.08981895& 21.09867477&	\textbf{20.77900505}\\
BA.4& \textbf{49.11387253}& 55.40925598&	66.97073364\\
BA.5& \textbf{806.7938232}& 852.3427124&	902.1066895\\
BQ.1& 134.2239838&	\textbf{134.1691589}&	134.2108307\\
C.37& \textbf{0.005054518}&	0.005466022&	0.005525291\\
Other& 7.389593601&	\textbf{5.299286366}&	6.80599308\\
P.1& 0.018325612&	0.020107888&	\textbf{0.15947794}9\\
P.3& 0.004451346&	\textbf{0.004076692}&	0.004459959\\
UNK& 0.286525726&	\textbf{0.022762863}&	0.292036057\\
XBB& 9.691458702&	\textbf{9.689269066}&	9.693496704\\
    \hline
	\end{tabular}
\end{table}

However, when we delve deeper onto each loss value for each variant, the loss value of BiLSTM tends to be smaller than the LSTM one. Therefore, we conclude that BiLSTM outperforms LSTM despite having the same frequency of minimum loss values. Therefore, the optimal configuration from the Multivariate test uses BiLSTM with hidden size 25 and layer size 4. 

Multivariate test results in our study show better results than the Univariate. Table \ref{table10} shows the MSE loss value of Univariate compared to Multivariate on LSTM. The majority of COVID-19 variants produce smaller loss values in Multivariate. In total, there are 13 variants with minimum loss values. This result is consistent with similar studies implementing Multivariate \cite{Yudistira2021} \cite{Said2021} \cite{Nick2020}. Multivariate can provide better results because more data is processed at one time, contrary to Univariate, in which computing one data at a time allows the model to train better and use data with more variety. 
\begin{table}[p]
	\centering
	\caption{Univariate vs Multivariate.}
	\label{table10}
	\begin{tabular}{@{}cccc@{}}
	\hline Variant& MSE Univariate& MSE Multivariate\\
    \hline
B.1.1.7& 0.285582066&  \textbf{0.030201122}\\
B.1.351& \textbf{0.000278501}& 0.000407222\\
B.1.427/B.1.429& 8.72087E-05& \textbf{7.28716E-06}\\
B.1.525& 9.04E-06& \textbf{4.07541E-05}\\
B.1616&	\textbf{0}& \textbf{0}\\
B.1.617.1& 9.93619E-05& \textbf{3.01427E-05}\\
B.1.617.2& \textbf{0.272779733}& 0.702166438\\
B.1.620& 0.000102891& \textbf{2.63529E-05}\\
B.1.621& 2.74123E-05& \textbf{1.16705E-05}\\
BA.1& \textbf{51.86838531}& 1181.40564\\
BA.2& 1069615.5& \textbf{374334}.625\\
BA.2.75& 445.1516724& \textbf{444.7804871}\\
BA.4& \textbf{2279.553711}& 2412.172363\\
BA.5& 680154.625& \textbf{650916.25}\\
BQ.1& \textbf{18005.15234}& 18016.07813\\
C.37& 2.77505E-05& \textbf{2.55481E-05}\\
Other&\textbf{4.598902225}& 54.60609055\\
P.1& 0.002071609& \textbf{0.000335828}\\
P.3& 0.000377038& \textbf{1.98145E-05}\\
UNK& \textbf{0.003363617}& 0.082096986\\
XBB& 561.8049927& \textbf{93.92437744}\\
    \hline
	\end{tabular}
\end{table}

\clearpage
\section{Conclusion}
The COVID-19 pandemic has become a serious global threat, with many variants emerging due to virus mutations. Since the first variant of COVID-19 was discovered in the UK until now, and the prediction of COVID-19 variants that have been identified is needed. This research aims to help governments and health organizations implement the right policies to deal with the pandemic. We are focusing on predicting confirmed cases of COVID-19 for each variant that has been identified. By utilizing ECDC's COVID-19 case dataset, we try to determine the optimal model configuration to be implemented by testing the number of hidden and layer sizes with Univariate and Multivariate forms. From the testing in our research, we concluded that in the Univariate test, the optimal configuration is using LSTM with the hidden size 25 and layer size 3. Multivariate uses LSTM with hidden size 25 and layer size 4. Both are proven by the frequency of the minimum loss value obtained from each COVID-19 variant.
\clearpage
 \bibliographystyle{elsarticle-num} 
 \bibliography{cas-refs}

\begin{thebibliography}{10}
\expandafter\ifx\csname url\endcsname\relax
  \def\url#1{\texttt{#1}}\fi
\expandafter\ifx\csname urlprefix\endcsname\relax\def\urlprefix{URL }\fi
\expandafter\ifx\csname href\endcsname\relax
  \def\href#1#2{#2} \def\path#1{#1}\fi

\bibitem{Gorbalenya2020}
A.~E. Gorbalenya, S.~C. Baker, R.~S. Baric, R.~J. de~Groot, C.~Drosten, B.~L.
  Gulyaeva, Anastasia A.~andHaagmans, C.~Lauber, A.~M. Leontovich, B.~W.
  Neuman, D.~Penzar, S.~Perlman, L.~L. Poon, D.~V. Samborskiy, I.~A. Sidorov,
  I.~Sola, J.~Ziebuhr, The species severe acute respiratory syndrome-related
  coronavirus: classifying 2019-ncov and naming it sars-cov-2, Nature
  Microbiology 5 (2020) 536--544.

\bibitem{WHO2021}
WHO, Who coronavirus disease (covid-19) dashboard, https://covid19.who.int/
  (accessed Feb. 03, 2021) (2021).

\bibitem{WHO2022}
WHO, Tracking sars-cov-2 variants,
  https://www.who.int/en/activities/tracking-SARS-CoV-2-variants/ (accessed
  Dec. 23, 2022) (2022).

\bibitem{Shahid2020}
F.~Shahid, A.~Zameer, M.~Muneeb, Predictions for covid-19 with deep learning
  models of lstm, gru and bi-lstm, Chaos, Solitons and Fractals 140 (2020)
  110212.

\bibitem{Parbat2020}
D.~Parbat, M.~Chakraborty, A python based support vector regression model for
  prediction of covid19 cases in india, Chaos, Solitons \& Fractals 138 (2020)
  109942.

\bibitem{Yudistira2020}
N.~Yudistira, Covid-19 growth prediction using multivariate long short term
  memory, arxiv 8 (2020) 1--9.

\bibitem{Yudistira2021}
N.~Yudistira, S.~Sumitro, N.~Bambang, R.~Alberth, F.~Nelly, Learning where to
  look for covid-19 growth: Multivariate analysis of covid-19 cases over time
  using explainable convolution–lstm, Applied Soft Computing 109 (2021)
  107469.

\bibitem{Chimmula2020}
V.~K.~R. Chimmula, L.~Zhang, Time series forecasting of covid-19 transmission
  in canada using lstm networks, Chaos, Solitons and Fractals 135 (2020).

\bibitem{Wang2020}
P.~Wang, X.~Zheng, G.~Ai, D.~Liu, B.~Zhu, Time series prediction for the
  epidemic trends of covid-19 using the improved lstm deep learning method:
  Case studies in russia, peru and iran, Chaos, Solitons and Fractals 140
  (2020) P10008.

\bibitem{Fillatre2021}
F.~Pierre, M.-J. Dufour, S.~Behillil, R.~Vatan, P.~Reusse, A.~Gabellec,
  N.~Velmans, C.~Montagne, S.~Coudret, E.~Droumaguet, V.~Merour, V.~Enouf,
  R.~Buzele, M.~Valence, E.~Guillotel, B.~Gagniere, A.~Baidaliuk, A.~Zhukova,
  M.~Tourdjman, N.~Massart, A new sars-cov-2 variant poorly detected by rt-pcr
  on nasopharyngeal samples, with high lethality: an observational study,
  Clinical microbiology and infection : the official publication of the
  European Society of Clinical Microbiology and Infectious Diseases 28 (10
  2021).
\newblock \href {https://doi.org/10.1016/j.cmi.2021.09.035}
  {\path{doi:10.1016/j.cmi.2021.09.035}}.

\bibitem{Dudas2021}
G.~Dudas, S.~L. Hong, B.~Potter, S.~Calvignac-Spencer, F.~S. Niatou-Singa,
  T.~B. Tombolomako, T.~Fuh-Neba, U.~Vickos, M.~Ulrich, F.~H. Leendertz,
  K.~Khan, A.~Watts, I.~Olendraite, J.~Snijder, K.~N. Wijnant, A.~M. Bonvin,
  P.~Martres, S.~Behillil, A.~Ayouba, M.~F. Maidadi, D.~M. Djomsi, C.~Godwe,
  C.~Butel, A.~{\v S}imaitis, M.~Gabrielaite, M.~Katenaite, R.~Norvilas,
  L.~Raugaite, R.~Jonikas, I.~Nasvytiene, {\v Z}.~{\v Z}emeckiene, D.~Ge{\v
  c}ys, K.~Tamu{\v s}auskaite, M.~Norkiene, E.~Vasiliunaite, D.~{\v Z}iogiene,
  A.~Timinskas, M.~{\v S}ukys, M.~{\v S}arauskas, G.~Alzbutas, D.~Juozapaite,
  D.~Naumovas, A.~Pautienius, A.~Vitkauskiene, R.~Ugenskiene, A.~Gedvilaite,
  D.~{\v C}ere{\v s}kevi{\v c}ius, V.~Lesauskaite, L.~{\v Z}emaitis, L.~Gri{\v
  s}kevi{\v c}ius, G.~Baele, Travel-driven emergence and spread of sars-cov-2
  lineage b.1.620 with multiple voc-like mutations and deletions in europe,
  medRxiv (2021).
\newblock \href {https://doi.org/10.1101/2021.05.04.21256637}
  {\path{doi:10.1101/2021.05.04.21256637}}.

\bibitem{Faria2021}
N.~R. Faria, I.~M. Claro, D.~Candido, L.~M. Franco, P.~S. Andrade, T.~M.
  Coletti, C.~A. Silva, F.~C. Sales, E.~R. Manuli, R.~S. Aguiar, et~al.,
  Genomic characterisation of an emergent sars-cov-2 lineage in manaus:
  preliminary findings, Virological 372 (2021) 815--821.

\bibitem{Francis2021}
F.~A. Tablizo, K.~M. Kim, C.~M. Lapid, M.~J.~R. Castro, M.~S.~L. Yangzon, B.~A.
  Maralit, M.~E.~C. Ayes, E.~M.~C. de~la Paz, A.~R.~D. Guzman, J.~M.~C. Yap,
  J.-H.~S. Llames, S.~M.~M. Araiza, K.~P. Punayan, I.~C.~A. Asin, C.~F.~B.
  Tambaoan, A.~L.~U. Chong, R.~P. S.~C. Karol Sophia Agape R.~Padilla, E.~K.~D.
  Morado, J.~G.~A. Dizon, R.~N.~M. Hao, A.~A. Zamora, D.~R. Pacial, J.~A.~R.
  Magalang, M.~Alejandria, C.~Carlos, A.~Ong-Lim, E.~M. Salvaña, J.~C.~M. John
  Q.~Wong, M.~R. Singh-Vergeire, C.~P. Saloma, Genome sequencing and analysis
  of an emergent sars-cov-2 variant characterized by multiple spike protein
  mutations detected from the central visayas region of the philippines,
  medRxiv 2021.03.03.21252812 (2021).
\newblock \href {https://doi.org/https://doi.org/10.1101/2021.03.03.21252812}
  {\path{doi:https://doi.org/10.1101/2021.03.03.21252812}}.

\bibitem{Rambaut2020}
A.~Rambaut, E.~C. Holmes, A.~O’Toole, V.~Hill, J.~T. McCrone, C.~Ruis,
  L.~du~Plessis, O.~G. Pybus, A dynamic nomenclature proposal for sars-cov-2
  lineages to assist genomic epidemiology, Nature Microbiology 11 (2020)
  1403–1407.

\bibitem{Yu2019}
Y.~Yong, S.~Xiaosheng, H.~Changhua, Z.~Jianxun, A review of recurrent neural
  networks: Lstm cells and network architectures, Neural computation 31 (2019)
  1235–1270.

\bibitem{Siami2019}
S.-N. Sima, T.~Neda, S.~N. Akbar, The performance of lstm and bilstm in
  forecasting time series, 2019 IEEE International Conference on Big Data (Big
  Data) (2019) 3285--3292.

\bibitem{Kingma2015}
D.~P. Kingma, J.~L. Ba, Adam: A method for stochastic optimization, 3rd
  International Conference on Learning Representations, ICLR 2015 - Conference
  Track Proceedings (2015) 1--15.

\bibitem{Chai2014}
T.~Chai, R.~R. Draxler, Root mean square error (rmse) or mean absolute error
  (mae)? -arguments against avoiding rmse in the literature, Geoscientific
  Model Development 7 (2014) 1247--1250.

\bibitem{Li2018}
S.~Li, W.~Li, C.~Cook, C.~Zhu, Y.~Gao, Independently recurrent neural network
  (indrnn): Building a longer and deeper rnn, Proceedings of the IEEE Computer
  Society Conference on Computer Vision and Pattern Recognition (2018)
  5457--5466.

\bibitem{Said2021}
A.~Said, A.~Erradi, H.~Aly, Predicting covid-19 cases using bidirectional lstm
  on multivariate time series, Environ Sci Pollut Res 28 (2021) 56043–56052.

\bibitem{Nick2020}
J.~Nick, M.~Max, Cluster-based dual evolution for multivariate time series:
  Analyzing covid-19, Chaos: An Interdisciplinary Journal of Nonlinear Science
  30 (2020).

\end{thebibliography}





\end{document}